\documentclass[11pt,a4paper]{article}
\usepackage{jheppub}
\usepackage{amsmath}
\usepackage[most]{tcolorbox}
\usepackage{dsfont}
\usepackage{ulem}
\usepackage{natbib}
\usepackage{xcolor}
\usepackage[hang,flushmargin]{footmisc}
\usepackage{tikz-cd}
\usepackage{enumitem}
\usepackage{amsfonts,amssymb, amscd,amsmath,latexsym,amsbsy,bm}
\usepackage{stmaryrd}
\usepackage{todonotes}
\usepackage{stackrel}
\usepackage{amsthm}

\usepackage{graphicx}

\title{\boldmath Algebraic Structures Behind the Yang-Baxterization Process}

\author{Cansu Özdemir$^{a}$, and Ilmar Gahramanov$^{b,c,d,e}$} 
\affiliation[a]{Cayyolu Doga Science and Technology High School, 06810 Cankaya/Ankara, Turkey}
\affiliation[b]{Department of Physics, Bogazici University, 34342 Bebek, Istanbul, Turkey}
\affiliation[c]{Steklov Mathematical Institute of Russian Academy of Sciences, 119991 Moscow, Russia}
\affiliation[d]{Department of Mathematics, Khazar University,  Mehseti St. 41, AZ1096, Baku, Azerbaijan}
\affiliation[e]{Institute of Radiation Problems ANAS, B.Vahabzade 9, AZ1143 Baku, Azerbaijan}

\emailAdd{cansuozdemir3.14@gmail.com}
\emailAdd{ilmar.gahramanov@boun.edu.tr}

\abstract{We review the Yang-Baxterization process of braid group representations. We discuss the corresponding $n$-CB algebras in the Yang-Baxterization process. We present diagrams of the relations for the $4$-CB algebras. These relations are illustrated using the isomorphism between the general free algebra generated by $\{1\}$, $\{E_i\}$, and $\{G_i\}$ and Kauffman's tangle algebra.}

\begin{document}

\maketitle

\section{Introduction}

Exactly solvable models play an important role in statistical physics, yet finding solvable models is not a simple task. The Yang-Baxter equation is a great tool for finding solvable models due to the fact that satisfying the Yang-Baxter equation is a sufficient condition for solvability \cite{Baxter:1982zz}. The Yang-Baxter equation was first introduced in \cite{Yang:1967bm} for the quantum many-particle system with $\delta$-potentials. Later it was used for the solution of the XYZ Heisenberg spin chain and coined its name in work \cite{Takhtajan:1979iv} where authors developed the quantum inverse scattering method. The study of this equation led to the invention of quantum groups \cite{Kulish:1981dli,Kulish:1981gi,Jimbo:1985vd, Jimbo:1985zk,Drinfeld:1985rx}. The Yang-Baxter equation appears in several areas of mathematical physics such as topological field theory, supersymmetric gauge theory, graph theory, number theory, knot theory, etc. We refer the interested reader to well-known review articles \cite{Thacker:1980ei,Kulish:1980ii, Wu:1992zzb} and recent works \cite{Gahramanov:2017ysd,Yamazaki:2018xbx,Gahramanov:2022qge,Isaev:2022mrc}.

The Yang-Baxter equation ensures that the transfer matrices commute\footnote{This is an analog of involutivity of the integrals of motion in classical integrability.} with each other. Therefore the problem of finding exactly solvable models with positive Boltzmann weights can be reduced to the problem of finding solutions to the Yang-Baxter equation. Finding solutions to this equation is still effort-taking. One way to generate solutions to the Yang-Baxter equation is using braid group representations, which is called the Yang-Baxterization procedure \cite{ge1991explicit}.\par

Braid groups and the Yang-Baxter equation are closely related, hence it is possible to obtain solutions to the Yang-Baxter equation from braid group representations (BGR) and obtain BGR's from Yang-Baxter equation solutions. In this paper, we will discuss the first case. For $n$ block case, there exists a corresponding algebra type that Yang-Baxter operators satisfy, and containing the representations of these algebras are usually the sufficient condition for a BGR with $n$ distinct eigenvalues to be Yang-Baxterizable.

\section{The Yang-Baxter equation}

In this section, the Yang-Baxter equation and Yang-Baxter operators will be introduced and the relations they satisfy will be discussed. Then the basic properties and symmetries of the models that satisfy the Yang-Baxter equation will be shown and the corresponding cases in the vertex model will be given and shown by diagrams.

\subsection{Yang-Baxter equation and Yang-Baxter operators}

Let $V$ be a complex vector space and $R(u)$ be a linear operator of $u \in \mathbb{C}$ which takes values in $End_{\mathbb{C}}(V \otimes V)$. The Yang-Baxter equation is the following  equation written in the operator form  \cite{jimbo1989introduction}
\begin{equation}
    R_{12}(u)R_{13}(u+v)R_{23}(v)=R_{23}(v)R_{13}(u+v)R_{12}(u)
\end{equation}
where $R_{ij}$ denotes the matrix on $V\otimes V \otimes V$ acting as $R(u)$ on the $i^{th}$ and $j^{th}$ component and as identity on the other. Here the variable $u$ is called the spectral parameter. Usually, it is assumed that the dimension of the vector space $V$ is finite.

Let $N=\dim(V)<\infty$. After choosing a basis, the $R$-matrix can be written in components
\begin{align}
    R(u)&=\sum R^{kl}_{ij}(u) E_{ik} \otimes E_{jl} \;,
\end{align}
where $E_{ij}=(\delta_{ia} \delta_{jb})_{a,b=1,\dots N}$. By writing it in this form the Yang-Baxter equation amounts to $N^6$ homogeneous equations for the $N^4$ unknowns $R^{kl}_{ij}$. Let $P \in End_{\mathbb{C}}(V \otimes V)$ denote the transposition
\begin{align}
    P x \otimes y = y \otimes x
\end{align}
Then if $R(u)$ has the property $R(0)=cP$ where $c \in \mathbb{C}$ is a constant. The relation $R(0)=cP$ is called the initial condition. Often the Yang-Baxter equation is written in terms of the matrix 
\begin{equation}
    \check{R}(u)=PR(u)
\end{equation}
Then for $m\geq 2$ a sequence of operators $R_1(u),R_2(u), \cdots, R_i(u), \cdots$ acting on $V^{\otimes m}$ can be defined such that
\begin{equation}
    \check{R}_i(u)= I^{(1)}\otimes I^{(2)}\otimes \cdots \otimes \check{R}(u) ^ {i \otimes i+1} \otimes I^{(i+2)} \otimes \cdots \otimes I^{(m)}
\end{equation}
where $I^{(j)}$ is the identity acting at the $j$-th position and $i=1,\cdots, m-1$. These operators satisfy the relations \cite{akutsu1989yang}:
\begin{align}
    \check{R}_i(u)\check{R}_j(v)=\check{R}_j(v)\check{R}_i(u) \; \textrm{      for } |i-j|\geq 2 \\
    \check{R}_i(u)\check{R}_{i+1}(u+v)\check{R}_i(v)=\check{R}_{i+1}(v)\check{R}_{i}(u+v)\check{R}_{i+1}(u)
\end{align}
The operators $\left\{ \check{R}_i(u) \right\}$ are called the Yang-Baxter operators. The Boltzmann weights satisfying the Yang-Baxter relations define a sequence of Yang-Baxter operators.

The order of the characteristic equation that Yang-Baxter operators satisfy in a model is called the block number of the model. For each $n$-block model, there is a corresponding $n$-CB algebra such that the model satisfies. The definitions and relations of these models are discussed in more detail in the section $n$-CB algebras.

Yang-Baxter equation takes different forms depending on the types of models under consideration. Yang-Baxter equation in terms of Boltzmann weights in the vertex model is the following:
\begin{equation}
    \sum_{\alpha, \beta, \gamma} S^{\beta q}_{\gamma r}(v) S^{\alpha p}_{k \gamma}(u+v)S^{i\alpha}_{j\beta}(u) = \sum_{\alpha, \beta, \gamma} S^{\alpha p}_{\beta q}(u) S^{i \alpha}_{ \gamma r}(u+v)S^{j\beta}_{k\gamma}(v)
\end{equation}
For vertex models, the Yang-Baxter operators $\left\{ \check{R}_i(u) \right\}$'s can be expressed as a sum of matrix tensor products:
\begin{equation}
    \check{R}_i(u)= \sum_{klmp}S^{km}_{lp}(u) \cdot I^{(1)}\otimes I^{(2)}\otimes \cdots \otimes E^{(i)}_{pk} \otimes E^{(i+1)}_{ml} \otimes I^{(i+2)} \otimes \cdots \otimes I^{(n)}
\end{equation}
where $I^{(j)}$ is the identity acting at the $j$-th position and $E_{jk}$ is a matrix such that $(E_{jk})_{pq}=\delta_{jp}\delta_{kq}$.\par
For the rest of this section, the vertex model will be used in order to illustrate the properties of the systems that satisfy the Yang-Baxter equation. \par
\begin{center}

\tikzset{every picture/.style={line width=0.75pt}} 

\begin{tikzpicture}[x=0.75pt,y=0.75pt,yscale=-1,xscale=1]

\draw    (189.64,182.89) -- (171.55,47.17) ;
\draw [shift={(171.28,45.19)}, rotate = 82.41] [color={rgb, 255:red, 0; green, 0; blue, 0 }  ][line width=0.75]    (10.93,-3.29) .. controls (6.95,-1.4) and (3.31,-0.3) .. (0,0) .. controls (3.31,0.3) and (6.95,1.4) .. (10.93,3.29)   ;
\draw    (275.32,112.51) -- (141.56,68.26) ;
\draw [shift={(139.66,67.63)}, rotate = 18.31] [color={rgb, 255:red, 0; green, 0; blue, 0 }  ][line width=0.75]    (10.93,-3.29) .. controls (6.95,-1.4) and (3.31,-0.3) .. (0,0) .. controls (3.31,0.3) and (6.95,1.4) .. (10.93,3.29)   ;
\draw    (158.02,174.73) -- (272.77,78.1) ;
\draw [shift={(274.3,76.81)}, rotate = 139.9] [color={rgb, 255:red, 0; green, 0; blue, 0 }  ][line width=0.75]    (10.93,-3.29) .. controls (6.95,-1.4) and (3.31,-0.3) .. (0,0) .. controls (3.31,0.3) and (6.95,1.4) .. (10.93,3.29)   ;
\draw  [draw opacity=0] (194,85.52) .. controls (194.29,86.69) and (194.39,87.92) .. (194.25,89.19) .. controls (193.64,95) and (188.42,99.22) .. (182.6,98.6) .. controls (181.03,98.43) and (179.57,97.93) .. (178.29,97.16) -- (183.72,88.07) -- cycle ; \draw   (194,85.52) .. controls (194.29,86.69) and (194.39,87.92) .. (194.25,89.19) .. controls (193.64,95) and (188.42,99.22) .. (182.6,98.6) .. controls (181.03,98.43) and (179.57,97.93) .. (178.29,97.16) ;  
\draw  [draw opacity=0] (188.06,171.07) .. controls (187.46,171.61) and (186.79,172.08) .. (186.06,172.48) .. controls (180.92,175.28) and (174.49,173.39) .. (171.69,168.25) .. controls (170.93,166.86) and (170.52,165.37) .. (170.42,163.88) -- (180.99,163.18) -- cycle ; \draw   (188.06,171.07) .. controls (187.46,171.61) and (186.79,172.08) .. (186.06,172.48) .. controls (180.92,175.28) and (174.49,173.39) .. (171.69,168.25) .. controls (170.93,166.86) and (170.52,165.37) .. (170.42,163.88) ;  
\draw  [draw opacity=0] (259.1,107.39) .. controls (258.13,112.38) and (254.4,116.62) .. (249.15,117.98) .. controls (242.94,119.59) and (236.62,116.64) .. (233.74,111.22) -- (245.74,104.81) -- cycle ; \draw   (259.1,107.39) .. controls (258.13,112.38) and (254.4,116.62) .. (249.15,117.98) .. controls (242.94,119.59) and (236.62,116.64) .. (233.74,111.22) ;

\draw    (469.97,182.08) -- (451.88,46.37) ;
\draw [shift={(451.61,44.39)}, rotate = 82.41] [color={rgb, 255:red, 0; green, 0; blue, 0 }  ][line width=0.75]    (10.93,-3.29) .. controls (6.95,-1.4) and (3.31,-0.3) .. (0,0) .. controls (3.31,0.3) and (6.95,1.4) .. (10.93,3.29)   ;
\draw    (500.57,161.68) -- (366.82,117.43) ;
\draw [shift={(364.92,116.8)}, rotate = 18.31] [color={rgb, 255:red, 0; green, 0; blue, 0 }  ][line width=0.75]    (10.93,-3.29) .. controls (6.95,-1.4) and (3.31,-0.3) .. (0,0) .. controls (3.31,0.3) and (6.95,1.4) .. (10.93,3.29)   ;
\draw    (371.04,153.52) -- (485.78,56.89) ;
\draw [shift={(487.31,55.6)}, rotate = 139.9] [color={rgb, 255:red, 0; green, 0; blue, 0 }  ][line width=0.75]    (10.93,-3.29) .. controls (6.95,-1.4) and (3.31,-0.3) .. (0,0) .. controls (3.31,0.3) and (6.95,1.4) .. (10.93,3.29)   ;
\draw  [draw opacity=0] (459.03,100.58) .. controls (458.43,101.11) and (457.76,101.59) .. (457.02,101.99) .. controls (451.89,104.79) and (445.46,102.9) .. (442.66,97.76) .. controls (441.9,96.37) and (441.48,94.88) .. (441.39,93.39) -- (451.96,92.69) -- cycle ; \draw   (459.03,100.58) .. controls (458.43,101.11) and (457.76,101.59) .. (457.02,101.99) .. controls (451.89,104.79) and (445.46,102.9) .. (442.66,97.76) .. controls (441.9,96.37) and (441.48,94.88) .. (441.39,93.39) ;  
\draw  [draw opacity=0] (483.4,155.69) .. controls (483.69,156.86) and (483.79,158.1) .. (483.65,159.36) .. controls (483.04,165.18) and (477.82,169.39) .. (472.01,168.78) .. controls (470.43,168.61) and (468.97,168.1) .. (467.69,167.34) -- (473.12,158.25) -- cycle ; \draw   (483.4,155.69) .. controls (483.69,156.86) and (483.79,158.1) .. (483.65,159.36) .. controls (483.04,165.18) and (477.82,169.39) .. (472.01,168.78) .. controls (470.43,168.61) and (468.97,168.1) .. (467.69,167.34) ;  
\draw  [draw opacity=0] (415.35,133.86) .. controls (414.38,138.85) and (410.64,143.09) .. (405.4,144.45) .. controls (399.19,146.06) and (392.87,143.11) .. (389.98,137.69) -- (401.99,131.28) -- cycle ; \draw   (415.35,133.86) .. controls (414.38,138.85) and (410.64,143.09) .. (405.4,144.45) .. controls (399.19,146.06) and (392.87,143.11) .. (389.98,137.69) ;

\draw (440.22,105.02) node [anchor=north west][inner sep=0.75pt]  [font=\normalsize]  {$u$};
\draw (477.77,168.52) node [anchor=north west][inner sep=0.75pt]  [font=\normalsize]  {$v$};
\draw (387.61,143.72) node [anchor=north west][inner sep=0.75pt]  [font=\normalsize]  {$u+v$};
\draw (168.57,175.51) node [anchor=north west][inner sep=0.75pt]  [font=\normalsize]  {$u$};
\draw (189.05,98.03) node [anchor=north west][inner sep=0.75pt]  [font=\normalsize]  {$v$};
\draw (230.68,117.93) node [anchor=north west][inner sep=0.75pt]  [font=\normalsize]  {$u+v$};
\draw (100.06,205.58) node [anchor=north west][inner sep=0.75pt]    {$\sum _{\alpha ,\beta ,\gamma } S_{\gamma r}^{\beta q}( v) \ S_{k\gamma }^{\alpha p}( u+v) S_{j\beta }^{i\alpha }( u)$};
\draw (306.23,105.23) node [anchor=north west][inner sep=0.75pt]  [font=\Large]  {$=$};
\draw (340.79,203.58) node [anchor=north west][inner sep=0.75pt]    {$\sum _{\alpha ,\beta ,\gamma } S_{\beta q}^{\alpha p}( u) \ S_{\gamma r}^{i\alpha }( u+v) S_{k\gamma }^{j\beta }( v)$};
\draw (165.32,26.33) node [anchor=north west][inner sep=0.75pt]    {$q$};
\draw (127.49,56.59) node [anchor=north west][inner sep=0.75pt]    {$r$};
\draw (277.14,63.32) node [anchor=north west][inner sep=0.75pt]    {$p$};
\draw (146.14,173.01) node [anchor=north west][inner sep=0.75pt]    {$i$};
\draw (187,182.01) node [anchor=north west][inner sep=0.75pt]    {$j$};
\draw (278.83,106.2) node [anchor=north west][inner sep=0.75pt]    {$k$};
\draw (167,110.4) node [anchor=north west][inner sep=0.75pt]    {$\beta $};
\draw (203.43,108.92) node [anchor=north west][inner sep=0.75pt]    {$\alpha $};
\draw (209.04,71.73) node [anchor=north west][inner sep=0.75pt]    {$\gamma $};
\draw (352.82,105.36) node [anchor=north west][inner sep=0.75pt]    {$r$};
\draw (446.98,25.48) node [anchor=north west][inner sep=0.75pt]    {$q$};
\draw (489.86,42.3) node [anchor=north west][inner sep=0.75pt]    {$p$};
\draw (360.38,148.24) node [anchor=north west][inner sep=0.75pt]    {$i$};
\draw (466.99,181.17) node [anchor=north west][inner sep=0.75pt]    {$j$};
\draw (503.32,155.81) node [anchor=north west][inner sep=0.75pt]    {$k$};
\draw (465.48,111.24) node [anchor=north west][inner sep=0.75pt]    {$\beta $};
\draw (413.63,92.95) node [anchor=north west][inner sep=0.75pt]    {$\alpha $};
\draw (431.01,119.65) node [anchor=north west][inner sep=0.75pt]    {$\gamma $};
\end{tikzpicture}

Figure 1: Yang-Baxter equation for vertex model.
\end{center}

\subsection{Properties and symmetries of models satisfying the Yang-Baxter equation}

Models that satisfy the Yang-Baxter equation often have some properties listed and explained below. Some of these properties are used in the process of Yang-Baxterization and finding representations of braid groups. Some of these properties can be used to obtain constraints in the Yang-Baxterization process.

\begin{itemize}

\item \textbf{Overall normalization} 

The Yang-Baxter relation is invariant under the following transformation.
\begin{align}
    \check{R}(u) \rightarrow N(u) \check{R}(u)
\end{align}
where $N(u)$ is an arbitrary function depending only on the spectral parameter. This demonstrates that the overall normalization of the Boltzmann weights can be chosen freely. 

\item \textbf{Standard initial condition} 

The standard initial condition states that there exists a case, where nothing changes; and when the spectral parameter is 0, this is the only case that exists.
\begin{equation}
    \check{R}(0) = C \cdot I
\end{equation}

\tikzset{every picture/.style={line width=0.75pt}} 
\begin{center}

\begin{tikzpicture}[x=0.75pt,y=0.75pt,yscale=-1,xscale=1]

\draw  [draw opacity=0] (201.9,144.6) .. controls (199.53,142.19) and (198,138.41) .. (198,134.17) .. controls (198,129.66) and (199.72,125.69) .. (202.34,123.31) -- (208,134.17) -- cycle ; \draw   (201.9,144.6) .. controls (199.53,142.19) and (198,138.41) .. (198,134.17) .. controls (198,129.66) and (199.72,125.69) .. (202.34,123.31) ;  
\draw    (312,133.83) -- (360,133.83)(312,136.83) -- (360,136.83) ;
\draw [shift={(368,135.33)}, rotate = 180] [color={rgb, 255:red, 0; green, 0; blue, 0 }  ][line width=0.75]    (10.93,-3.29) .. controls (6.95,-1.4) and (3.31,-0.3) .. (0,0) .. controls (3.31,0.3) and (6.95,1.4) .. (10.93,3.29)   ;
\draw [shift={(304,135.33)}, rotate = 0] [color={rgb, 255:red, 0; green, 0; blue, 0 }  ][line width=0.75]    (10.93,-3.29) .. controls (6.95,-1.4) and (3.31,-0.3) .. (0,0) .. controls (3.31,0.3) and (6.95,1.4) .. (10.93,3.29)   ;

\draw (158,90.4) node [anchor=north west][inner sep=0.75pt]    {$i$};
\draw (157,162.4) node [anchor=north west][inner sep=0.75pt]    {$j$};
\draw (266,90.4) node [anchor=north west][inner sep=0.75pt]    {$l$};
\draw (265,163.4) node [anchor=north west][inner sep=0.75pt]    {$k$};
\draw (185,126.4) node [anchor=north west][inner sep=0.75pt]    {$0$};
\draw (397,91.4) node [anchor=north west][inner sep=0.75pt]    {$i$};
\draw (396,163.4) node [anchor=north west][inner sep=0.75pt]    {$j$};
\draw (505,91.4) node [anchor=north west][inner sep=0.75pt]    {$i$};
\draw (504,164.4) node [anchor=north west][inner sep=0.75pt]    {$j$};
\draw (193,184.4) node [anchor=north west][inner sep=0.75pt]    {$S^{ik}_{jl}(0)$};
\draw (426,184.4) node [anchor=north west][inner sep=0.75pt]    {$C \cdot \delta_{il}\delta_{jk} $};
\draw    (173,104.14) -- (260.35,163.73) ;
\draw [shift={(262,164.86)}, rotate = 214.3] [color={rgb, 255:red, 0; green, 0; blue, 0 }  ][line width=0.75]    (10.93,-3.29) .. controls (6.95,-1.4) and (3.31,-0.3) .. (0,0) .. controls (3.31,0.3) and (6.95,1.4) .. (10.93,3.29)   ;
\draw    (172,164.06) -- (261.33,105.05) ;
\draw [shift={(263,103.94)}, rotate = 146.55] [color={rgb, 255:red, 0; green, 0; blue, 0 }  ][line width=0.75]    (10.93,-3.29) .. controls (6.95,-1.4) and (3.31,-0.3) .. (0,0) .. controls (3.31,0.3) and (6.95,1.4) .. (10.93,3.29)   ;
\draw    (412,106.72) .. controls (442.21,135.36) and (471.77,135.87) .. (500.68,108.25) ;
\draw [shift={(502,106.97)}, rotate = 135.41] [color={rgb, 255:red, 0; green, 0; blue, 0 }  ][line width=0.75]    (10.93,-3.29) .. controls (6.95,-1.4) and (3.31,-0.3) .. (0,0) .. controls (3.31,0.3) and (6.95,1.4) .. (10.93,3.29)   ;
\draw    (411,163.54) .. controls (441.21,135.83) and (470.77,135.6) .. (499.68,162.87) ;
\draw [shift={(501,164.13)}, rotate = 224.21] [color={rgb, 255:red, 0; green, 0; blue, 0 }  ][line width=0.75]    (10.93,-3.29) .. controls (6.95,-1.4) and (3.31,-0.3) .. (0,0) .. controls (3.31,0.3) and (6.95,1.4) .. (10.93,3.29)   ;

\end{tikzpicture}

Figure 2: Diagrammatic representation of the standard initial condition for the vertex model.
\end{center}

Here, C is a constant depending on the overall normalization of the Boltzmann weights.

\item \textbf{Unitarity condition or the first inversion relation}

The unitarity condition (first inversion relation) has the following form
\begin{equation}
   \check{R}(-u)\check{R}(u)=\rho(u)\rho(-u)I \;.
\end{equation}
The function $\rho(u)$ is called \textit{unitarity function}. We can always choose a normalization\footnote{Using the overall normalization property.} where $\rho(u) \equiv 1$. Let us choose $u=0$ to deduce the relation between the spectral parameter $u$ and $C$.\begin{align}
   \check{R}(0)\check{R}(0)&=\rho(0)\rho(0)I\\
    C\cdot C&= \rho(0)^2
\end{align}
From here one can choose a normalization of $\rho(u)$ such that $\rho(0)=C$.

For the vertex model, the unitarity condition can be written as (visualized in Figure 3.)
\begin{equation}
    \sum_{p, q} S^{ql}_{pk}(-u)S^{ip}_{jq}(u)=\rho(u)\rho(-u)\delta_{ik}\delta_{jl} \;.
\end{equation}

\tikzset{every picture/.style={line width=0.75pt}} 
\begin{center}
\begin{tikzpicture}[x=0.75pt,y=0.75pt,yscale=-1,xscale=1]

\draw  [draw opacity=0] (246.9,151.6) .. controls (244.53,149.19) and (243,145.41) .. (243,141.17) .. controls (243,136.66) and (244.72,132.69) .. (247.34,130.31) -- (253,141.17) -- cycle ; \draw   (246.9,151.6) .. controls (244.53,149.19) and (243,145.41) .. (243,141.17) .. controls (243,136.66) and (244.72,132.69) .. (247.34,130.31) ;  
\draw  [draw opacity=0] (397.9,151.6) .. controls (395.53,149.19) and (394,145.41) .. (394,141.17) .. controls (394,136.66) and (395.72,132.69) .. (398.34,130.31) -- (404,141.17) -- cycle ; \draw   (397.9,151.6) .. controls (395.53,149.19) and (394,145.41) .. (394,141.17) .. controls (394,136.66) and (395.72,132.69) .. (398.34,130.31) ;  

\draw (203,97.4) node [anchor=north west][inner sep=0.75pt]    {$i$};
\draw (202,169.4) node [anchor=north west][inner sep=0.75pt]    {$j$};
\draw (311,97.4) node [anchor=north west][inner sep=0.75pt]    {$q$};
\draw (310,170.4) node [anchor=north west][inner sep=0.75pt]    {$p$};
\draw (230,130.4) node [anchor=north west][inner sep=0.75pt]    {$u$};
\draw (354,97.4) node [anchor=north west][inner sep=0.75pt]    {$q$};
\draw (353,169.4) node [anchor=north west][inner sep=0.75pt]    {$p$};
\draw (462,97.4) node [anchor=north west][inner sep=0.75pt]    {$k$};
\draw (461,170.4) node [anchor=north west][inner sep=0.75pt]    {$l$};
\draw (371,130.4) node [anchor=north west][inner sep=0.75pt]    {$-u$};
\draw (290,205.4) node [anchor=north west][inner sep=0.75pt]    {$S^{ql}_{pk}(-u)S^{ip}_{jq}(u)$};
\draw    (218,111.14) -- (305.35,170.73) ;
\draw [shift={(307,171.86)}, rotate = 214.3] [color={rgb, 255:red, 0; green, 0; blue, 0 }  ][line width=0.75]    (10.93,-3.29) .. controls (6.95,-1.4) and (3.31,-0.3) .. (0,0) .. controls (3.31,0.3) and (6.95,1.4) .. (10.93,3.29)   ;
\draw    (217,171.06) -- (306.33,112.05) ;
\draw [shift={(308,110.94)}, rotate = 146.55] [color={rgb, 255:red, 0; green, 0; blue, 0 }  ][line width=0.75]    (10.93,-3.29) .. controls (6.95,-1.4) and (3.31,-0.3) .. (0,0) .. controls (3.31,0.3) and (6.95,1.4) .. (10.93,3.29)   ;
\draw    (369,111.14) -- (456.35,170.73) ;
\draw [shift={(458,171.86)}, rotate = 214.3] [color={rgb, 255:red, 0; green, 0; blue, 0 }  ][line width=0.75]    (10.93,-3.29) .. controls (6.95,-1.4) and (3.31,-0.3) .. (0,0) .. controls (3.31,0.3) and (6.95,1.4) .. (10.93,3.29)   ;
\draw    (368,171.06) -- (457.33,112.05) ;
\draw [shift={(459,110.94)}, rotate = 146.55] [color={rgb, 255:red, 0; green, 0; blue, 0 }  ][line width=0.75]    (10.93,-3.29) .. controls (6.95,-1.4) and (3.31,-0.3) .. (0,0) .. controls (3.31,0.3) and (6.95,1.4) .. (10.93,3.29)   ;
\draw    (326,105) -- (349,105) ;
\draw [shift={(351,105)}, rotate = 180] [color={rgb, 255:red, 0; green, 0; blue, 0 }  ][line width=0.75]    (10.93,-3.29) .. controls (6.95,-1.4) and (3.31,-0.3) .. (0,0) .. controls (3.31,0.3) and (6.95,1.4) .. (10.93,3.29)   ;
\draw    (325,177.79) -- (348,177.26) ;
\draw [shift={(350,177.21)}, rotate = 178.67] [color={rgb, 255:red, 0; green, 0; blue, 0 }  ][line width=0.75]    (10.93,-3.29) .. controls (6.95,-1.4) and (3.31,-0.3) .. (0,0) .. controls (3.31,0.3) and (6.95,1.4) .. (10.93,3.29)   ;

\end{tikzpicture}

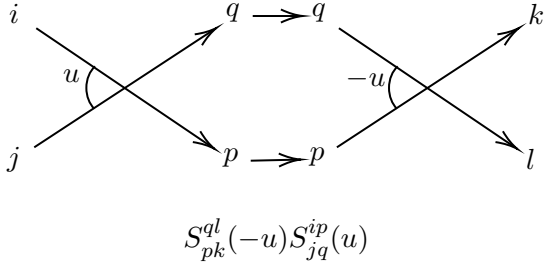
Figure 3: Diagrammatic representation of the first inversion relation for the vertex model.
\end{center}

\item \textbf{Crossing symmetry}

Some models have a symmetry that enables them to be interpreted from two directions. The models that have this property are said to have crossing symmetry. For vertex models that have crossing symmetry, this relation can be written as:
\begin{equation}
    S^{ik}_{jl}(u) = \left[ \frac{r(i)r(l)}{r(j)r(k)} \right]^{\frac{1}{2}} F(u) \cdot S^{jl}_{\bar{k}\bar{i}}(\lambda- u)
\end{equation}
This case is represented in diagrams in Figure 4. Here, $\lambda$ is called the crossing point of the spectral parameter or \textit{crossing parameter} for short. For the vertex model, the bar-operation on the indices usually denotes charge conjugation $\bar{i}=-i$, yet different interpretations of the bar-operation are also possible. \par
The factors $\left\{ r(i)\right\}$ are called \textit{crossing multipliers}. In relation with $r(\bar{i})=\frac{1}{r(i)}$. Crossing multipliers are added to the equation to prevent any deterioration in the physical interpretation of the Boltzmann weights. \par
The function $F(u)$ also depends on the overall normalization and always satisfies $F(u)F(\lambda - u) = 1$. We can normalize the Boltzmann weights so that $F(u) \equiv 1$.\par
\begin{center}
\tikzset{every picture/.style={line width=0.75pt}} 

\begin{tikzpicture}[x=0.75pt,y=0.75pt,yscale=-1,xscale=1]

\draw  [draw opacity=0] (444.66,142.77) .. controls (442.16,148.01) and (436.29,151.7) .. (429.45,151.72) .. controls (421.73,151.74) and (415.24,147.1) .. (413.38,140.81) -- (429.4,137.1) -- cycle ; \draw   (444.66,142.77) .. controls (442.16,148.01) and (436.29,151.7) .. (429.45,151.72) .. controls (421.73,151.74) and (415.24,147.1) .. (413.38,140.81) ;  
\draw    (300,131.33) -- (360,131.33) ;
\draw [shift={(362,131.33)}, rotate = 180] [color={rgb, 255:red, 0; green, 0; blue, 0 }  ][line width=0.75]    (10.93,-3.29) .. controls (6.95,-1.4) and (3.31,-0.3) .. (0,0) .. controls (3.31,0.3) and (6.95,1.4) .. (10.93,3.29)   ;
\draw [shift={(298,131.33)}, rotate = 0] [color={rgb, 255:red, 0; green, 0; blue, 0 }  ][line width=0.75]    (10.93,-3.29) .. controls (6.95,-1.4) and (3.31,-0.3) .. (0,0) .. controls (3.31,0.3) and (6.95,1.4) .. (10.93,3.29)   ;
\draw  [draw opacity=0] (201.82,142.91) .. controls (198.32,140.52) and (196,136.3) .. (196,131.5) .. controls (196,126.45) and (198.56,122.05) .. (202.36,119.74) -- (208.5,131.5) -- cycle ; \draw   (201.82,142.91) .. controls (198.32,140.52) and (196,136.3) .. (196,131.5) .. controls (196,126.45) and (198.56,122.05) .. (202.36,119.74) ;  

\draw (365,72.73) node [anchor=north west][inner sep=0.75pt]    {$\overline{i}$};
\draw (362,170.73) node [anchor=north west][inner sep=0.75pt]    {$j$};
\draw (481,74.73) node [anchor=north west][inner sep=0.75pt]    {$l$};
\draw (481,167.73) node [anchor=north west][inner sep=0.75pt]    {$\overline{k}$};
\draw (409,156.73) node [anchor=north west][inner sep=0.75pt]    {$\lambda -u$};
\draw (155,75.4) node [anchor=north west][inner sep=0.75pt]    {$i$};
\draw (155,171.4) node [anchor=north west][inner sep=0.75pt]    {$j$};
\draw (264,74.4) node [anchor=north west][inner sep=0.75pt]    {$l$};
\draw (263,171.4) node [anchor=north west][inner sep=0.75pt]    {$k$};
\draw (180,123.4) node [anchor=north west][inner sep=0.75pt]    {$u$};
\draw (195,193.4) node [anchor=north west][inner sep=0.75pt]    {$S^{ik}_{jl}(u)$};
\draw (338,190.4) node [anchor=north west][inner sep=0.75pt]    {$\left[ \frac{r(i)r(l)}{r(j)r(k)} \right]^{\frac{1}{2}} F(u) \cdot S^{jl}_{\bar{k}\bar{i}}(\lambda- u)$};
\draw    (381.55,90.97) -- (478,169.96) ;
\draw [shift={(380,89.7)}, rotate = 39.32] [color={rgb, 255:red, 0; green, 0; blue, 0 }  ][line width=0.75]    (10.93,-3.29) .. controls (6.95,-1.4) and (3.31,-0.3) .. (0,0) .. controls (3.31,0.3) and (6.95,1.4) .. (10.93,3.29)   ;
\draw    (377,171.07) -- (476.44,90.85) ;
\draw [shift={(478,89.59)}, rotate = 141.11] [color={rgb, 255:red, 0; green, 0; blue, 0 }  ][line width=0.75]    (10.93,-3.29) .. controls (6.95,-1.4) and (3.31,-0.3) .. (0,0) .. controls (3.31,0.3) and (6.95,1.4) .. (10.93,3.29)   ;
\draw    (170,91) -- (258.51,169.67) ;
\draw [shift={(260,171)}, rotate = 221.63] [color={rgb, 255:red, 0; green, 0; blue, 0 }  ][line width=0.75]    (10.93,-3.29) .. controls (6.95,-1.4) and (3.31,-0.3) .. (0,0) .. controls (3.31,0.3) and (6.95,1.4) .. (10.93,3.29)   ;
\draw    (170,170.99) -- (259.51,91.34) ;
\draw [shift={(261,90.01)}, rotate = 138.33] [color={rgb, 255:red, 0; green, 0; blue, 0 }  ][line width=0.75]    (10.93,-3.29) .. controls (6.95,-1.4) and (3.31,-0.3) .. (0,0) .. controls (3.31,0.3) and (6.95,1.4) .. (10.93,3.29)   ;

\end{tikzpicture}

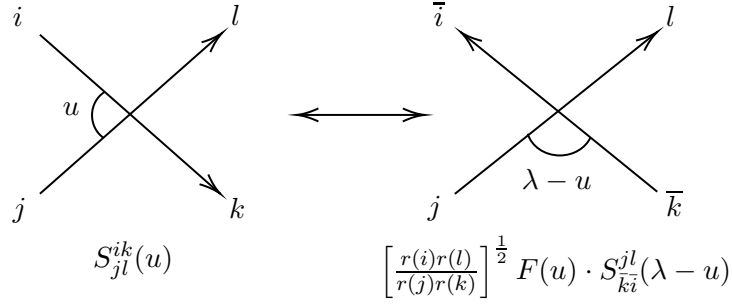
Figure 4: Diagrammatic representation of the crossing symmetry for the vertex model.
\end{center}

\item \textbf{Second inversion relation} 

Some models satisfy a relation called the second inversion relation. The second inversion relation for the vertex models is as follows:
\begin{equation}
    \sum_{p, q} S^{kp}_{ql}(\lambda + u) S^{jq}_{pi}(\lambda - u) \frac{r(q)r(p)}{r(j)r(k)} = \rho (u) \rho (-u) \delta_{ik}\delta_{jl}
\end{equation}
\begin{center}
\tikzset{every picture/.style={line width=0.75pt}} 

\begin{tikzpicture}[x=0.75pt,y=0.75pt,yscale=-1,xscale=1]

\draw  [draw opacity=0] (403.9,81.4) .. controls (401.53,78.99) and (400,75.21) .. (400,70.97) .. controls (400,66.46) and (401.72,62.49) .. (404.34,60.11) -- (410,70.97) -- cycle ; \draw   (403.9,81.4) .. controls (401.53,78.99) and (400,75.21) .. (400,70.97) .. controls (400,66.46) and (401.72,62.49) .. (404.34,60.11) ;  
\draw  [draw opacity=0] (206.9,81.4) .. controls (204.53,78.99) and (203,75.21) .. (203,70.97) .. controls (203,66.46) and (204.72,62.49) .. (207.34,60.11) -- (213,70.97) -- cycle ; \draw   (206.9,81.4) .. controls (204.53,78.99) and (203,75.21) .. (203,70.97) .. controls (203,66.46) and (204.72,62.49) .. (207.34,60.11) ;  

\draw (360,27.2) node [anchor=north west][inner sep=0.75pt]    {$k$};
\draw (361,100.2) node [anchor=north west][inner sep=0.75pt]    {$q$};
\draw (468,27.2) node [anchor=north west][inner sep=0.75pt]    {$l$};
\draw (467,100.2) node [anchor=north west][inner sep=0.75pt]    {$p$};
\draw (357,60.2) node [anchor=north west][inner sep=0.75pt]    {$\lambda +u$};
\draw (163,27.2) node [anchor=north west][inner sep=0.75pt]    {$j$};
\draw (162,99.2) node [anchor=north west][inner sep=0.75pt]    {$p$};
\draw (271,27.2) node [anchor=north west][inner sep=0.75pt]    {$i$};
\draw (270,100.2) node [anchor=north west][inner sep=0.75pt]    {$q$};
\draw (161,60.2) node [anchor=north west][inner sep=0.75pt]    {$\lambda -u$};
\draw (187,122.4) node [anchor=north west][inner sep=0.75pt]    {$S_{pi}^{jq}( \lambda -u)$};
\draw (382,124.4) node [anchor=north west][inner sep=0.75pt]    {$S_{ql}^{kp}( \lambda +u) \ $};
\draw    (375,40.94) -- (462.35,100.53) ;
\draw [shift={(464,101.66)}, rotate = 214.3] [color={rgb, 255:red, 0; green, 0; blue, 0 }  ][line width=0.75]    (10.93,-3.29) .. controls (6.95,-1.4) and (3.31,-0.3) .. (0,0) .. controls (3.31,0.3) and (6.95,1.4) .. (10.93,3.29)   ;
\draw    (376,101.66) -- (463.35,42.07) ;
\draw [shift={(465,40.94)}, rotate = 145.7] [color={rgb, 255:red, 0; green, 0; blue, 0 }  ][line width=0.75]    (10.93,-3.29) .. controls (6.95,-1.4) and (3.31,-0.3) .. (0,0) .. controls (3.31,0.3) and (6.95,1.4) .. (10.93,3.29)   ;
\draw    (178,40.94) -- (265.35,100.53) ;
\draw [shift={(267,101.66)}, rotate = 214.3] [color={rgb, 255:red, 0; green, 0; blue, 0 }  ][line width=0.75]    (10.93,-3.29) .. controls (6.95,-1.4) and (3.31,-0.3) .. (0,0) .. controls (3.31,0.3) and (6.95,1.4) .. (10.93,3.29)   ;
\draw    (177,100.86) -- (266.33,41.85) ;
\draw [shift={(268,40.74)}, rotate = 146.55] [color={rgb, 255:red, 0; green, 0; blue, 0 }  ][line width=0.75]    (10.93,-3.29) .. controls (6.95,-1.4) and (3.31,-0.3) .. (0,0) .. controls (3.31,0.3) and (6.95,1.4) .. (10.93,3.29)   ;
\end{tikzpicture}

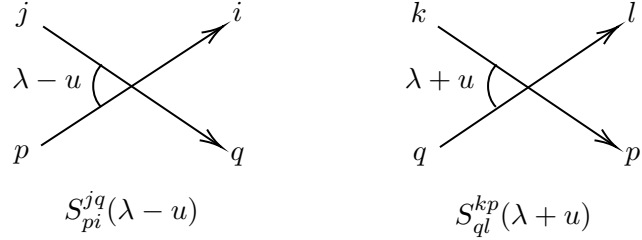
Figure 5: Diagrammatic representation of the second inversion relation for the vertex model.
\end{center}
For models with crossing symmetry, the second inversion relation is equivalent to the unitary relation. Which is shown below:
\allowdisplaybreaks
\begin{align}
    &S^{kp}_{ql}(\lambda + u) = \left[ \frac{r(k)r(l)}{r(p)r(q)} \right]^{\frac{1}{2}} F(\lambda + u) S^{ql}_{\bar{k}\bar{p}}(\lambda- (\lambda + u))  \nonumber \\
    &S^{jq}_{pi}(\lambda - u) = \left[ \frac{r(j)r(i)}{r(p)r(q)} \right]^{\frac{1}{2}} F(\lambda - u) S^{pi}_{\bar{j}\bar{q}}(\lambda- (\lambda - u))  \nonumber \\
    &\begin{aligned}
    \sum_{p, q}  S^{ql}_{\bar{k}\bar{p}}(- u)) S^{pi}_{\bar{j}\bar{q}}(u)) \left[\frac{r(k)r(l)}{r(p)r(q)} \right]^{\frac{1}{2}} \left[\frac{r(j)r(i)}{r(p)r(q)} \right]^{\frac{1}{2}}  \left[\frac{r(q)r(p)}{r(j)r(k)} \right] &F(\lambda + u) F(\lambda - u)\\
    &=\rho (u) \rho (-u) \delta_{ik}\delta_{jl} 
    \end{aligned}  \nonumber \\
    &\sum_{p, q} S^{ql}_{\bar{k}\bar{p}}(- u) S^{pi}_{\bar{j}\bar{q}}(u) \left[\frac{r(k)r(j)}{r(p)r(q)} \right] \left[\frac{r(q)r(p)}{r(j)r(k)} \right] F(\lambda + u) F(\lambda - u) =\rho (u) \rho (-u) \delta_{ik}\delta_{jl} \nonumber \\
    & \sum_{p, q} S^{ql}_{\bar{k}\bar{p}}(- u) S^{pi}_{\bar{j}\bar{q}}(u) F(\lambda + u) F(\lambda - u) =\rho (u) \rho (-u) \delta_{ik}\delta_{jl}
\end{align}

\begin{center}
\tikzset{every picture/.style={line width=0.75pt}} 

\begin{tikzpicture}[x=0.75pt,y=0.75pt,yscale=-1,xscale=1]

\draw  [draw opacity=0] (205.9,96.4) .. controls (203.53,93.99) and (202,90.21) .. (202,85.97) .. controls (202,81.46) and (203.72,77.49) .. (206.34,75.11) -- (212,85.97) -- cycle ; \draw   (205.9,96.4) .. controls (203.53,93.99) and (202,90.21) .. (202,85.97) .. controls (202,81.46) and (203.72,77.49) .. (206.34,75.11) ;  
\draw  [draw opacity=0] (399.9,96.4) .. controls (397.53,93.99) and (396,90.21) .. (396,85.97) .. controls (396,81.46) and (397.72,77.49) .. (400.34,75.11) -- (406,85.97) -- cycle ; \draw   (399.9,96.4) .. controls (397.53,93.99) and (396,90.21) .. (396,85.97) .. controls (396,81.46) and (397.72,77.49) .. (400.34,75.11) ;  

\draw (196,132.4) node [anchor=north west][inner sep=0.75pt]    {$\ S_{\overline{j}\overline{q}}^{pi}( u)$};
\draw (161,41.2) node [anchor=north west][inner sep=0.75pt]    {$p$};
\draw (162,114.2) node [anchor=north west][inner sep=0.75pt]    {$\overline{j}$};
\draw (269,41.2) node [anchor=north west][inner sep=0.75pt]    {$\overline{q}$};
\draw (268,114.2) node [anchor=north west][inner sep=0.75pt]    {$i$};
\draw (184,74.2) node [anchor=north west][inner sep=0.75pt]    {$u$};
\draw (353,42.2) node [anchor=north west][inner sep=0.75pt]    {$q$};
\draw (352,114.2) node [anchor=north west][inner sep=0.75pt]    {$\overline{k}$};
\draw (467,42.2) node [anchor=north west][inner sep=0.75pt]    {$\overline{p}$};
\draw (466,115.2) node [anchor=north west][inner sep=0.75pt]    {$l$};
\draw (372,75.2) node [anchor=north west][inner sep=0.75pt]    {$-u$};
\draw (385,132.4) node [anchor=north west][inner sep=0.75pt]    {$S_{\overline{k}\overline{p}}^{ql}( -\ u)$};
\draw    (176,54.94) -- (263.35,114.53) ;
\draw [shift={(265,115.66)}, rotate = 214.3] [color={rgb, 255:red, 0; green, 0; blue, 0 }  ][line width=0.75]    (10.93,-3.29) .. controls (6.95,-1.4) and (3.31,-0.3) .. (0,0) .. controls (3.31,0.3) and (6.95,1.4) .. (10.93,3.29)   ;
\draw    (177,117.49) -- (264.36,56.26) ;
\draw [shift={(266,55.11)}, rotate = 144.97] [color={rgb, 255:red, 0; green, 0; blue, 0 }  ][line width=0.75]    (10.93,-3.29) .. controls (6.95,-1.4) and (3.31,-0.3) .. (0,0) .. controls (3.31,0.3) and (6.95,1.4) .. (10.93,3.29)   ;
\draw    (368,55.61) -- (461.32,115.9) ;
\draw [shift={(463,116.99)}, rotate = 212.86] [color={rgb, 255:red, 0; green, 0; blue, 0 }  ][line width=0.75]    (10.93,-3.29) .. controls (6.95,-1.4) and (3.31,-0.3) .. (0,0) .. controls (3.31,0.3) and (6.95,1.4) .. (10.93,3.29)   ;
\draw    (367,118.01) -- (462.32,56.67) ;
\draw [shift={(464,55.59)}, rotate = 147.24] [color={rgb, 255:red, 0; green, 0; blue, 0 }  ][line width=0.75]    (10.93,-3.29) .. controls (6.95,-1.4) and (3.31,-0.3) .. (0,0) .. controls (3.31,0.3) and (6.95,1.4) .. (10.93,3.29)   ;

\end{tikzpicture}

Figure 6
\end{center}

From here it has been shown that for models with crossing symmetry, the second inversion relation is equivalent to the unitarity relation. There are, however, solvable models satisfying the Yang-Baxter relation \textit{without} the crossing symmetry. Nevertheless, those models satisfy the second inversion relation.

\item \textbf{CPT invariances or reflection symmetry} 

Solvable models that satisfy the Yang-Baxter relation have three additional symmetries, called the CPT invariances or reflection symmetries. For the vertex model, these symmetries are as follows:
\begin{align}
  & \; \; = S^{\bar{i}\bar{k}}_{\bar{j}\bar{l}}(u)  \nonumber  \\
    S^{ik}_{jl}(u) \; \; & \; \;= S^{jl}_{ik}(u)  \nonumber  \\
  & \; \; = S^{ki}_{lj}(u)
\end{align}
The diagrammatic representation of these symmetries is given in Figure 7.

\begin{center}

\tikzset{every picture/.style={line width=0.75pt}} 

\begin{tikzpicture}[x=0.75pt,y=0.75pt,yscale=-1,xscale=1]

\draw  [draw opacity=0] (134.1,209.95) .. controls (132.11,207.93) and (130.83,204.76) .. (130.83,201.19) .. controls (130.83,197.4) and (132.27,194.06) .. (134.47,192.07) -- (139.23,201.19) -- cycle ; \draw   (134.1,209.95) .. controls (132.11,207.93) and (130.83,204.76) .. (130.83,201.19) .. controls (130.83,197.4) and (132.27,194.06) .. (134.47,192.07) ;  

\draw  [draw opacity=0] (369.05,80.95) .. controls (367.05,78.93) and (365.77,75.76) .. (365.77,72.19) .. controls (365.77,68.4) and (367.21,65.06) .. (369.42,63.07) -- (374.18,72.19) -- cycle ; \draw   (369.05,80.95) .. controls (367.05,78.93) and (365.77,75.76) .. (365.77,72.19) .. controls (365.77,68.4) and (367.21,65.06) .. (369.42,63.07) ;  

\draw  [draw opacity=0] (371.57,213.74) .. controls (369.57,211.72) and (368.29,208.54) .. (368.29,204.97) .. controls (368.29,201.19) and (369.74,197.85) .. (371.94,195.85) -- (376.7,204.97) -- cycle ; \draw   (371.57,213.74) .. controls (369.57,211.72) and (368.29,208.54) .. (368.29,204.97) .. controls (368.29,201.19) and (369.74,197.85) .. (371.94,195.85) ;  

\draw  [draw opacity=0] (373.25,349.05) .. controls (371.26,347.02) and (369.97,343.85) .. (369.97,340.28) .. controls (369.97,336.5) and (371.42,333.15) .. (373.62,331.16) -- (378.38,340.28) -- cycle ; \draw   (373.25,349.05) .. controls (371.26,347.02) and (369.97,343.85) .. (369.97,340.28) .. controls (369.97,336.5) and (371.42,333.15) .. (373.62,331.16) ;

\draw (126.56,236.67) node [anchor=north west][inner sep=0.75pt]    {$S_{jl}^{ik}( u)$};
\draw (358.58,105.51) node [anchor=north west][inner sep=0.75pt]    {$S_{\overline{j}\overline{k}}^{\overline{i}\overline{k}}( u)$};
\draw (333.72,166.98) node [anchor=north west][inner sep=0.75pt]    {$j$};
\draw (332.88,227.49) node [anchor=north west][inner sep=0.75pt]    {$i$};
\draw (424.48,166.98) node [anchor=north west][inner sep=0.75pt]    {$k$};
\draw (423.64,228.33) node [anchor=north west][inner sep=0.75pt]    {$l$};
\draw (355.57,194.71) node [anchor=north west][inner sep=0.75pt]    {$u$};
\draw (332.04,32.19) node [anchor=north west][inner sep=0.75pt]    {$\overline{i}$};
\draw (331.2,93.54) node [anchor=north west][inner sep=0.75pt]    {$\overline{j}$};
\draw (422.8,32.19) node [anchor=north west][inner sep=0.75pt]    {$\overline{l}$};
\draw (421.96,94.38) node [anchor=north west][inner sep=0.75pt]    {$\overline{k}$};
\draw (353.89,62.77) node [anchor=north west][inner sep=0.75pt]    {$u$};
\draw (96.25,163.19) node [anchor=north west][inner sep=0.75pt]    {$i$};
\draw (95.41,223.7) node [anchor=north west][inner sep=0.75pt]    {$j$};
\draw (187.02,163.19) node [anchor=north west][inner sep=0.75pt]    {$l$};
\draw (186.18,224.54) node [anchor=north west][inner sep=0.75pt]    {$k$};
\draw (118.1,190.93) node [anchor=north west][inner sep=0.75pt]    {$u$};
\draw (362.34,240.3) node [anchor=north west][inner sep=0.75pt]    {$S_{ik}^{jl}( u)$};
\draw (335.4,302.29) node [anchor=north west][inner sep=0.75pt]    {$k$};
\draw (334.56,362.8) node [anchor=north west][inner sep=0.75pt]    {$l$};
\draw (426.16,302.29) node [anchor=north west][inner sep=0.75pt]    {$j$};
\draw (425.32,363.64) node [anchor=north west][inner sep=0.75pt]    {$i$};
\draw (357.25,330.02) node [anchor=north west][inner sep=0.75pt]    {$u$};
\draw (364.02,371.56) node [anchor=north west][inner sep=0.75pt]    {$S_{lj}^{ki}( u)$};
\draw (261.4,59.82) node [anchor=north west][inner sep=0.75pt]  [font=\Large]  {$=$};
\draw (261.4,189.79) node [anchor=north west][inner sep=0.75pt]  [font=\Large]  {$=$};
\draw (262.46,324.26) node [anchor=north west][inner sep=0.75pt]  [font=\Large]  {$=$};
\draw (463,63) node [anchor=north west][inner sep=0.75pt]   [align=left] {C invariance};
\draw (464,197) node [anchor=north west][inner sep=0.75pt]   [align=left] {P invariance};
\draw (465,331) node [anchor=north west][inner sep=0.75pt]   [align=left] {T invariance};
\draw    (111.25,176.93) -- (181.53,224.88) ;
\draw [shift={(183.18,226)}, rotate = 214.3] [color={rgb, 255:red, 0; green, 0; blue, 0 }  ][line width=0.75]    (10.93,-3.29) .. controls (6.95,-1.4) and (3.31,-0.3) .. (0,0) .. controls (3.31,0.3) and (6.95,1.4) .. (10.93,3.29)   ;
\draw    (110.41,225.36) -- (182.35,177.84) ;
\draw [shift={(184.02,176.74)}, rotate = 146.55] [color={rgb, 255:red, 0; green, 0; blue, 0 }  ][line width=0.75]    (10.93,-3.29) .. controls (6.95,-1.4) and (3.31,-0.3) .. (0,0) .. controls (3.31,0.3) and (6.95,1.4) .. (10.93,3.29)   ;
\draw    (347.04,48.02) -- (417.32,96.62) ;
\draw [shift={(418.96,97.76)}, rotate = 214.67] [color={rgb, 255:red, 0; green, 0; blue, 0 }  ][line width=0.75]    (10.93,-3.29) .. controls (6.95,-1.4) and (3.31,-0.3) .. (0,0) .. controls (3.31,0.3) and (6.95,1.4) .. (10.93,3.29)   ;
\draw    (346.2,97.12) -- (418.14,48.93) ;
\draw [shift={(419.8,47.82)}, rotate = 146.19] [color={rgb, 255:red, 0; green, 0; blue, 0 }  ][line width=0.75]    (10.93,-3.29) .. controls (6.95,-1.4) and (3.31,-0.3) .. (0,0) .. controls (3.31,0.3) and (6.95,1.4) .. (10.93,3.29)   ;
\draw    (348.72,180.72) -- (418.99,228.66) ;
\draw [shift={(420.64,229.79)}, rotate = 214.3] [color={rgb, 255:red, 0; green, 0; blue, 0 }  ][line width=0.75]    (10.93,-3.29) .. controls (6.95,-1.4) and (3.31,-0.3) .. (0,0) .. controls (3.31,0.3) and (6.95,1.4) .. (10.93,3.29)   ;
\draw    (347.88,229.14) -- (419.81,181.63) ;
\draw [shift={(421.48,180.52)}, rotate = 146.55] [color={rgb, 255:red, 0; green, 0; blue, 0 }  ][line width=0.75]    (10.93,-3.29) .. controls (6.95,-1.4) and (3.31,-0.3) .. (0,0) .. controls (3.31,0.3) and (6.95,1.4) .. (10.93,3.29)   ;
\draw    (350.4,316.03) -- (420.67,363.97) ;
\draw [shift={(422.32,365.1)}, rotate = 214.3] [color={rgb, 255:red, 0; green, 0; blue, 0 }  ][line width=0.75]    (10.93,-3.29) .. controls (6.95,-1.4) and (3.31,-0.3) .. (0,0) .. controls (3.31,0.3) and (6.95,1.4) .. (10.93,3.29)   ;
\draw    (349.56,364.45) -- (421.49,316.93) ;
\draw [shift={(423.16,315.83)}, rotate = 146.55] [color={rgb, 255:red, 0; green, 0; blue, 0 }  ][line width=0.75]    (10.93,-3.29) .. controls (6.95,-1.4) and (3.31,-0.3) .. (0,0) .. controls (3.31,0.3) and (6.95,1.4) .. (10.93,3.29)   ;

\end{tikzpicture}

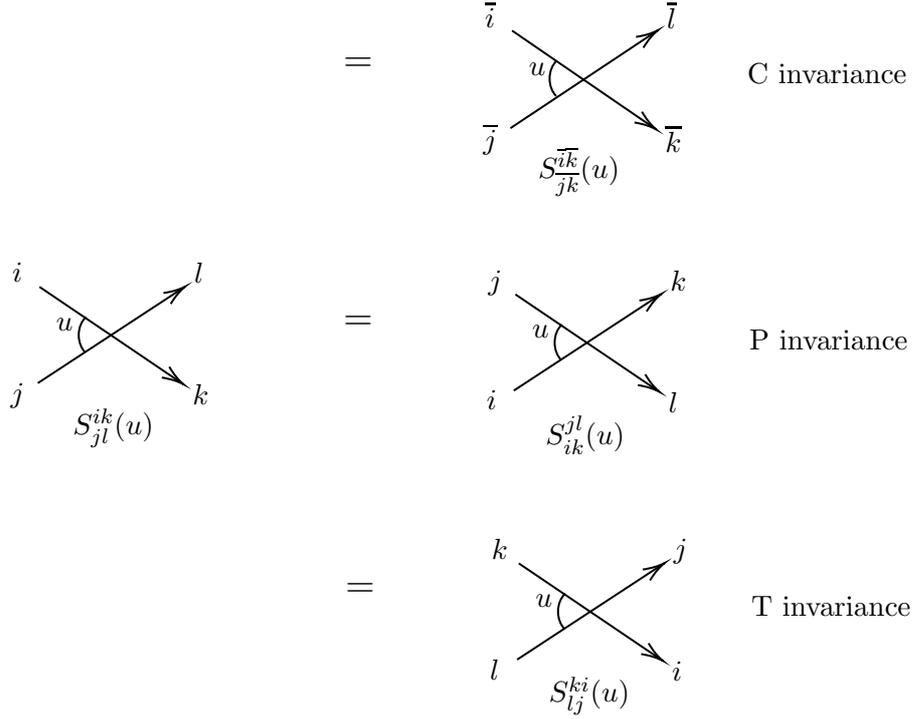
Figure 7: diagrammatic representations of CPT invariances for the vertex model.
\end{center}

\item \textbf{Charge conservation condition} 

Some models satisfy a relation called \textit{charge conservation condition}, which can be expressed as the following for the vertex model.
\begin{equation}
    S^{ik}_{jl}(u) = 0 \textrm{ unless } i+j=k+l
\end{equation}
Remark that not all of the solvable vertex models satisfy the charge conservation condition, however, the existence of a nontrivial link polynomials associated with the model requires the charge conservation condition \cite{akutsu1989yang}. 

\end{itemize}

\section{Braid group and BGR's}

Braid group of $n$ strings, denoted by \(B_n\), is the group generated by \(\{b_i\}\) and \(\{b_i^{-1}\}\) where \(i=1, \ 2, \cdots n-1\). The generators of the braid group can be visualized as shown in Figure 8.

\begin{center}
\tikzset{every picture/.style={line width=0.75pt}} 
\begin{tikzpicture}[x=0.75pt,y=0.75pt,yscale=-1,xscale=1]

\draw    (167.12,69.26) -- (202.77,119.91) ;
\draw    (181.17,99.62) -- (167.53,119.34) ;
\draw    (202.37,69.26) -- (188.52,90.66) ;

\draw    (238.75,69.47) -- (238.32,120.9) ;
\draw    (288.6,69.47) -- (288.17,120.9) ;
\draw    (84.47,69.09) -- (84.04,119.59) ;
\draw    (134.32,68.16) -- (133.89,119.59) ;

\draw    (547.42,70.06) -- (547.51,122.23) ;
\draw    (393.15,70.26) -- (393.24,122.44) ;
\draw    (443.72,69.76) -- (443.81,121.94) ;
\draw    (490.72,91.14) -- (474.97,69.72) ;
\draw    (475.06,121.91) -- (512.24,69.96) ;
\draw    (497.47,100.39) -- (512.32,121.53) ;

\draw    (596.84,70.06) -- (596.87,88.83) -- (596.93,122.23) ;

\draw (163.58,51.1) node [anchor=north west][inner sep=0.75pt]  [font=\small]  {$i$};
\draw (189.08,50.16) node [anchor=north west][inner sep=0.75pt]  [font=\small]  {$i+1$};
\draw (121.98,51.1) node [anchor=north west][inner sep=0.75pt]  [font=\small]  {$i-1$};
\draw (57.03,83.54) node [anchor=north west][inner sep=0.75pt]    {$b_{i}$};
\draw (225.44,50.16) node [anchor=north west][inner sep=0.75pt]  [font=\small]  {$i+2$};
\draw (100.67,79.85) node [anchor=north west][inner sep=0.75pt]    {$...$};
\draw (255.89,82.09) node [anchor=north west][inner sep=0.75pt]    {$...$};
\draw (79.67,51.1) node [anchor=north west][inner sep=0.75pt]  [font=\small]  {$1$};
\draw (283.86,50.16) node [anchor=north west][inner sep=0.75pt]  [font=\small]  {$n$};
\draw (336.35,81.7) node [anchor=north west][inner sep=0.75pt]    {$b_{i}^{-1}$};
\draw (472.2,52.03) node [anchor=north west][inner sep=0.75pt]  [font=\small]  {$i$};
\draw (497.69,51.1) node [anchor=north west][inner sep=0.75pt]  [font=\small]  {$i+1$};
\draw (430.6,52.03) node [anchor=north west][inner sep=0.75pt]  [font=\small]  {$i-1$};
\draw (534.06,51.1) node [anchor=north west][inner sep=0.75pt]  [font=\small]  {$i+2$};
\draw (388.28,52.03) node [anchor=north west][inner sep=0.75pt]  [font=\small]  {$1$};
\draw (592.47,51.1) node [anchor=north west][inner sep=0.75pt]  [font=\small]  {$n$};
\draw (564.16,82.81) node [anchor=north west][inner sep=0.75pt]  [rotate=-359.43]  {$...$};
\draw (409.89,83.01) node [anchor=north west][inner sep=0.75pt]  [rotate=-359.43]  {$...$};

\end{tikzpicture}
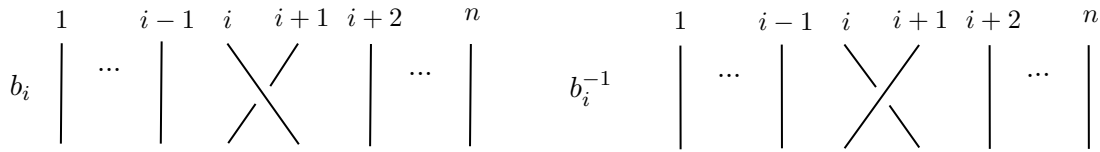
\\ Figure 8: Diagrammatic representation of the generators of the braid group.
\end{center}

The braid group was first introduced by Artin in 1925 \cite{artin1925theorie}. In this work, the defining relations between the generators of the braid group was shown. These relations are known as Artin relations, which are given below.
\begin{align}
    &b_i b_j = b_j b_i; \ |i-j| \geq 2  \nonumber \\
    &b_i b_{i+1} b_i= b_{i+1} b_i b_{i+1}
\end{align}
The diagrammatic representation of the Artin relations are given in Figure 9. 
\begin{center}

\tikzset{every picture/.style={line width=0.75pt}} 

\begin{tikzpicture}[x=0.75pt,y=0.75pt,yscale=-1,xscale=1]

\draw    (252,42) -- (280.32,82.25) ;
\draw    (263.16,66.12) -- (252.33,81.79) ;
\draw    (280,42) -- (269,59) ;

\draw    (225,81) -- (253.32,121.25) ;
\draw    (236.16,105.12) -- (225.33,120.79) ;
\draw    (253,81) -- (242,98) ;

\draw    (252,120) -- (280.32,160.25) ;
\draw    (263.16,144.12) -- (252.33,159.79) ;
\draw    (280,120) -- (269,137) ;

\draw    (225.33,120.79) -- (225,160) ;
\draw    (280.32,82.25) -- (280,120) ;
\draw    (225.33,41.79) -- (225,81) ;
\draw    (88,42) -- (116.32,82.25) ;
\draw    (99.16,66.12) -- (88.33,81.79) ;
\draw    (116,42) -- (105,59) ;

\draw    (115,80) -- (143.32,120.25) ;
\draw    (126.16,104.12) -- (115.33,119.79) ;
\draw    (143,80) -- (132,97) ;

\draw    (88,119) -- (116.32,159.25) ;
\draw    (99.16,143.12) -- (88.33,158.79) ;
\draw    (116,119) -- (105,136) ;

\draw    (88.33,81.79) -- (88,119) ;
\draw    (143.33,41.79) -- (143,80) ;
\draw    (143.32,120.25) -- (143,159.45) ;

\draw    (361,64) -- (389.32,104.25) ;
\draw    (372.16,88.12) -- (361.33,103.79) ;
\draw    (389,64) -- (378,81) ;

\draw    (389.32,104.25) -- (389,160.45) ;
\draw    (361.33,103.79) -- (361,160) ;
\draw    (433.24,140.11) -- (403.98,100.54) ;
\draw    (421.51,116.26) -- (431.98,100.34) ;
\draw    (405.24,140.77) -- (415.84,123.51) ;

\draw    (403.98,100.54) -- (403.56,73.19) -- (403.41,42.33) ;
\draw    (431.98,100.34) -- (432.41,42.13) ;
\draw    (544,63) -- (572.32,103.25) ;
\draw    (555.16,87.12) -- (544.33,102.79) ;
\draw    (572,63) -- (561,80) ;

\draw    (572.32,103.25) -- (572,140.45) ;
\draw    (544.33,102.79) -- (544,140) ;
\draw    (529.24,140.11) -- (499.98,100.54) ;
\draw    (517.51,116.26) -- (527.98,100.34) ;
\draw    (501.24,140.77) -- (511.84,123.51) ;

\draw    (499.98,100.54) -- (499.41,42.33) ;
\draw    (527.98,100.34) -- (528.41,43.13) ;
\draw    (389.33,42.79) -- (389,64) ;
\draw    (361.33,42.79) -- (361,64) ;
\draw    (405.24,140.77) -- (404.92,160.98) ;
\draw    (433.24,140.11) -- (432.91,160.32) ;
\draw    (544.33,42.79) -- (544,63) ;
\draw    (572.33,42.79) -- (572,63) ;
\draw    (544,140) -- (543.67,160.21) ;
\draw    (572,140.45) -- (571.67,160.66) ;
\draw    (501.24,140.77) -- (500.92,160.98) ;
\draw    (529.24,140.11) -- (528.91,160.32) ;

\draw (452,89.4) node [anchor=north west][inner sep=0.75pt]    {$=$};
\draw (384,168.07) node [anchor=north west][inner sep=0.75pt]    {$b_{i} b_{j}$};
\draw (524,169.07) node [anchor=north west][inner sep=0.75pt]    {$b_{j} b_{i}$};
\draw (84,21.4) node [anchor=north west][inner sep=0.75pt]    {$i$};
\draw (223,21.4) node [anchor=north west][inner sep=0.75pt]    {$i$};
\draw (358,21.4) node [anchor=north west][inner sep=0.75pt]    {$i$};
\draw (496,19.4) node [anchor=north west][inner sep=0.75pt]    {$i$};
\draw (399,19.4) node [anchor=north west][inner sep=0.75pt]    {$j$};
\draw (540,19.4) node [anchor=north west][inner sep=0.75pt]    {$j$};
\draw (170,91.4) node [anchor=north west][inner sep=0.75pt]    {$=$};
\draw (90,168.07) node [anchor=north west][inner sep=0.75pt]    {$b_{i} b_{i+1} b_{i}$};
\draw (222,168.07) node [anchor=north west][inner sep=0.75pt]    {$b_{i+1} b_{i} b_{i+1}$};

\end{tikzpicture}

Figure 9: Diagrammatic representation of the Artin relations.
\end{center}
\subsection{Braids and Yang-Baxter operators}
There is a close relationship between the braid group and the Yang-Baxter operators. The relations Yang-Baxter operators satisfy are very similar to Artin relations. The Yang-Baxter operators additionally have spectral parameters. In order to make a transition from one to the other, we either need to find a way to get rid of the spectral parameters or add spectral parameters.  By figuring out a way for making Yang-Baxter operators' spectral parameters independent, one can obtain BGR's using Boltzmann weights. One way to do this is to set all of the spectral parameters equal to each other.
\begin{equation}
    u=u+v=v
\end{equation}
Some of the ways this can be done are setting them all to 0, taking the limit as $u$ goes to infinity, or setting them all equal to the crossing parameter of the model examined. By doing this, one can obtain representations of the braid group. For the rest of this paper, let us generally assume that by taking the limit $\lim_{u\rightarrow u_0} u$ we obtain a BGR from Yang-Baxter operators.\par
BGR's can be categorized with respect to the number of eigenvalues a $B$ matrix of a representation of the braid group has. This classification will be used in the transition between Yang-Baxter operators and BGR's. As mentioned before, there are different algebraic structures contained in $n$-block models, and the corresponding property in BGR's is the number of distinct eigenvalues that BGR has. For notational convenience, let $n$ be number of the distinct eigenvalues $\lambda_i$ ($i=1,2,\cdots, n$) $B$ has. 

\subsection{Braids and knots}
In 1930, Alexander stated a theorem known as ``Alexander's Theorem'', which states that by gluing together the strands starting at $i^{th}$ place and ending at $i^{th}$ place, the braid can be ``closed" to obtain a knot, and vice versa, every knot can be represented as a closed braid \cite{alexander1923lemma}. However, in this paper it was also stated that there is no unique representation of a knot by a closed braid. This led to the question ``Which braid closures give isotopic knots?'' Markov's theorem \cite{akutsu1989yang} states that the closures of the braids that can be transformed into each other by the Markov moves give isotopic knots. The Markov moves are shown in Figure 10.
\begin{align}
    &AB \longrightarrow BA \nonumber \\
    &A \longrightarrow Ab_n \ , A \longrightarrow Ab_n^{-1}
\end{align}
\begin{center}
\tikzset{every picture/.style={line width=0.75pt}} 

\begin{tikzpicture}[x=0.75pt,y=0.75pt,yscale=-1,xscale=1]

\draw   (166.65,78.64) -- (103.33,78.99) -- (103.15,47.17) -- (166.47,46.82) -- cycle ;
\draw    (106.6,26.32) -- (106.71,47.8) ;
\draw    (134.59,26) -- (134.7,47.48) ;
\draw    (162.29,26.15) -- (162.4,47.63) ;

\draw    (106.6,79.32) -- (106.71,100.8) ;
\draw    (134.59,79) -- (134.7,100.48) ;
\draw    (162.29,79.15) -- (162.4,100.63) ;

\draw   (166.65,132.64) -- (103.33,132.99) -- (103.15,101.17) -- (166.47,100.82) -- cycle ;
\draw    (106.6,133.32) -- (106.71,154.8) ;
\draw    (134.59,133) -- (134.7,154.48) ;
\draw    (162.29,133.15) -- (162.4,154.63) ;

\draw   (279.65,78.64) -- (216.33,78.99) -- (216.15,47.17) -- (279.47,46.82) -- cycle ;
\draw    (219.6,26.32) -- (219.71,47.8) ;
\draw    (247.59,26) -- (247.7,47.48) ;
\draw    (275.29,26.15) -- (275.4,47.63) ;

\draw    (219.6,79.32) -- (219.71,100.8) ;
\draw    (247.59,79) -- (247.7,100.48) ;
\draw    (275.29,79.15) -- (275.4,100.63) ;

\draw   (279.65,132.64) -- (216.33,132.99) -- (216.15,101.17) -- (279.47,100.82) -- cycle ;
\draw    (219.6,133.32) -- (219.71,154.8) ;
\draw    (247.59,133) -- (247.7,154.48) ;
\draw    (275.29,133.15) -- (275.4,154.63) ;

\draw    (175.6,90) -- (206.6,90) ;
\draw [shift={(208.6,90)}, rotate = 180] [color={rgb, 255:red, 0; green, 0; blue, 0 }  ][line width=0.75]    (10.93,-3.29) .. controls (6.95,-1.4) and (3.31,-0.3) .. (0,0) .. controls (3.31,0.3) and (6.95,1.4) .. (10.93,3.29)   ;
\draw [shift={(173.6,90)}, rotate = 0] [color={rgb, 255:red, 0; green, 0; blue, 0 }  ][line width=0.75]    (10.93,-3.29) .. controls (6.95,-1.4) and (3.31,-0.3) .. (0,0) .. controls (3.31,0.3) and (6.95,1.4) .. (10.93,3.29)   ;

\draw   (424.65,108.64) -- (361.33,108.99) -- (361.15,77.17) -- (424.47,76.82) -- cycle ;
\draw    (364.6,28.71) -- (364.71,77) ;
\draw    (392.59,28) -- (392.7,76.29) ;
\draw    (420.29,28.34) -- (420.4,76.62) ;

\draw    (364.6,109.7) -- (364.71,156.8) ;
\draw    (392.59,109) -- (392.7,156.1) ;
\draw    (420.29,109.33) -- (420.4,156.43) ;

\draw    (434.6,92) -- (465.6,92) ;
\draw [shift={(467.6,92)}, rotate = 180] [color={rgb, 255:red, 0; green, 0; blue, 0 }  ][line width=0.75]    (10.93,-3.29) .. controls (6.95,-1.4) and (3.31,-0.3) .. (0,0) .. controls (3.31,0.3) and (6.95,1.4) .. (10.93,3.29)   ;
\draw [shift={(432.6,92)}, rotate = 0] [color={rgb, 255:red, 0; green, 0; blue, 0 }  ][line width=0.75]    (10.93,-3.29) .. controls (6.95,-1.4) and (3.31,-0.3) .. (0,0) .. controls (3.31,0.3) and (6.95,1.4) .. (10.93,3.29)   ;
\draw   (539.65,108.64) -- (476.33,108.99) -- (476.15,77.17) -- (539.47,76.82) -- cycle ;
\draw    (479.6,28.71) -- (479.71,77) ;
\draw    (507.59,28) -- (507.7,76.29) ;
\draw    (535.29,28.34) -- (535.4,76.62) ;

\draw    (479.6,109.7) -- (479.71,156.8) ;
\draw    (507.59,109) -- (507.7,156.1) ;
\draw    (535.29,109.33) -- (536,116) ;
\draw    (564.59,28) -- (564,116) ;
\draw    (536,116) -- (564.32,156.25) ;
\draw    (547.16,140.12) -- (536.33,155.79) ;
\draw    (564,116) -- (553,133) ;

\draw (122.34,159.4) node [anchor=north west][inner sep=0.75pt]    {$AB$};
\draw (235.98,159.4) node [anchor=north west][inner sep=0.75pt]    {$BA$};
\draw (127,54.4) node [anchor=north west][inner sep=0.75pt]    {$A$};
\draw (240,108.4) node [anchor=north west][inner sep=0.75pt]    {$A$};
\draw (129,109.4) node [anchor=north west][inner sep=0.75pt]    {$B$};
\draw (242,54.4) node [anchor=north west][inner sep=0.75pt]    {$B$};
\draw (385.34,161.4) node [anchor=north west][inner sep=0.75pt]    {$A$};
\draw (505.98,161.4) node [anchor=north west][inner sep=0.75pt]    {$Ab_{n}$};
\draw (385,83.4) node [anchor=north west][inner sep=0.75pt]    {$A$};
\draw (500,83.4) node [anchor=north west][inner sep=0.75pt]    {$A$};
\draw (530,7.4) node [anchor=north west][inner sep=0.75pt]    {$n$};
\draw (415,8.4) node [anchor=north west][inner sep=0.75pt]    {$n$};
\end{tikzpicture}

Figure 10: Markov moves.
\end{center}

Every closed braid gives rise to links, and braids which can be converted to each other by Markov moves and Artin relations give rise to isotopic links, therefore braid representations form a natural tool for constructing link invariants. By finding a quantity on a braid group representation, which is invariant under the Markov moves, a link invariant can be constructed. Using the relation between Yang-Baxter operators and Artin relations, solvable models that satisfy the Yang-Baxter equation can be used to obtain braid group representation and hence be used for finding knot invariants. In Akutsu, Deuguchi, and Wadati's work \cite{akutsu1989yang} it is emphasised that the theory of solvable models provides a powerful background for constructing new link polynomials.

\section{Yang-Baxterization}

The similarity between the relations that Yang-Baxter operators satisfy and the Artin relations leads to the idea of Yang-Baxterization, which can be defined as the process of obtaining solutions to the Yang-Baxter equation using braid group representations. The idea of "Baxterization" was first introduced by Jones \cite{ge1991explicit}. The methods of Yang-Baxterization have been studied by many authors \cite{ge1991explicit}. In \cite{cheng1991yang} it was shown that the inclusion of the Temperley-Lieb algebra is a sufficient condition for obtaining Yang-Baxter equation solutions for the 2-block model. Then, in \cite{li1993yang}, methods of Yang-Baxterization using the Birman-Murakami-Wenzl (BMW) algebra were discussed. There are many methods of Yang-Baxterization \cite{ge1991explicit}. In this paper, only the methods using the $n$-CB algebras are discussed. 

For a BGR with $n$ distinct eigenvalues to be Yang-Baxterizable, it must contain representations of the corresponding algebras, generally called $n$-CB algebras. If this condition is satisfied, then a BGR can be Yang-Baxterized with an appropriate choice of a function of the spectral parameters. The function of spectral parameters used in Yang-Baxterization is not unique, but depending on the properties and symmetries of models satisfying the Yang-Baxter equation, some constraints on the function can be found. In this section, the constraints on the function of the spectral parameter will be obtained using properties of the models that satisfy the Yang-Baxter equation and the ansatz for obtaining BGR's from Yang-Baxter operators, then the $n$-CB algebras for $n=2,3,4$ will be discussed.

\subsection{Spectral parameter function}

In this section, how the Yang-Baxter operators will be expressed using the representation of the generators of the braid group and the constraints on the function of spectral parameter which will be used in Yang-Baxterization will be obtained \cite{li1993yang} .\par
Let B have $n$ distinct eigenvalues. Then $B$ obeys an $n^{th}$ order reduction relation
\begin{equation}
    \prod_{i=1}^{n} (B - \lambda_i I) =0
\end{equation}
Then $B$ can be written as a sum of projector matrices.
\begin{equation}
    B = \sum_{i=1}^{n} \lambda_i P_i
\end{equation}
Where $P_i$ is the projector matrix onto the subspace of $\lambda_i$ and $I = \sum_{i=1}^{n} P_i$. $P_i$'s can be uniquely written as
\begin{equation}
    P_i = \prod_{i\neq j} \frac{S-\lambda_j}{\lambda_i-\lambda_j}
\end{equation} 
Using $\lambda_i$'s and $P_i$'s, it is possible to construct solutions to the Yang-Baxter equation. It is important to note that for different permutations of the eigenvalues of the BGR, it is possible to get different solutions of the Yang-Baxter equation from the same BGR. The most basic prescription for Yang-Baxterization is 
\begin{equation}
    \check{R}(u) = \sum_{i=1} ^{n} \Lambda_i(u) P_i
\end{equation}
The main problem is to determine the functions $\Lambda_i(u)$. Now substituting (38) in the basic properties described in the first section, we can obtain some constraints that  $\Lambda_i(u)$'s need to satisfy. \par
Consider the process of obtaining a BGR from the Yang-Baxter operators.
\begin{equation}
    B=\lim_{u \to u_0} \check{R}(u) = \sum_{i=1} ^{n} \Lambda_i(u_0) P_i= \sum_{i=1}^{n} \lambda_i P_i
\end{equation}
From here it can be seen that $\Lambda_i(u_0)=\lambda_i$ should be satisfied. From the initial condition, it is clear that $\Lambda_i(0)=\Lambda_j(0)$ for every $i$ and $j$. Lastly, using the first inversion relation, $\Lambda_i(u)\Lambda_i(-u)=\Lambda_j(u)\Lambda_j(-u)$ for any $u$. We now have some relations that $\Lambda_i(u)$'s needs to satisfy. These constraints on $\Lambda_i(u)$'s will be very useful in the process of Yang-Baxterization.\par
Without much difficulty, it can be found that \cite{li1993yang}
\begin{equation} \label{specpar}
    \Lambda_i (u) = \prod_{j=1} ^{i-1} \left( \frac{\lambda_j}{\lambda_{j+1}} f_j(u) +1 \right) \prod_{k=i} ^{n-1} \left( f_k(u) + \frac{\lambda_k}{\lambda_{k+1}}\right) \lambda_n
\end{equation}
By writing out $\Lambda_i(u)$'s like this, we can separate the function that is dependent on the spectral parameter and the eigenvalues of the $B$ matrix. Again, we will try to find some constraints on $f_l(u)$'s by using the properties mentioned in the first section. 
\begin{equation}
    \check{R}(u_0) = \sum_{i}^{n}\left[ \prod_{j=1} ^{i-1} \left( \frac{\lambda_j}{\lambda_{j+1}} f_j(u_0) +1 \right) \prod_{k=i} ^{n-1} \left( f_k(u_0) + \frac{\lambda_k}{\lambda_{k+1}}\right) \lambda_n \right] P_i
\end{equation}
From (41) it is clearly seen that $f_l(u_0)=0$ should be satisfied. And from the first inversion relation 
\begin{equation}
    \check{R}(0) = \sum_{i}^{n}\left[ \prod_{j=1} ^{i-1} \left( \frac{\lambda_j}{\lambda_{j+1}} f_j(0) +1 \right) \prod_{k=i} ^{n-1} \left( f_k(0) + \frac{\lambda_k}{\lambda_{k+1}}\right) \lambda_n \right] P_i =C\cdot I
\end{equation}$f_l(0)=1$, since  $\sum_{i=1}^{n} P_i = I $. Now we have some tools we can use for Yang-Baxterization. There are multiple ways to construct solutions to the Yang-Baxter equation after this point, as an example, let us investigate the case where $f_l(u)$'s are index independent, namely, $f_l(u)=y(u)$ for every $l$ value. The equation (\ref{specpar}) then becomes
\begin{equation}
    \Lambda_i (u) = \prod_{j=1} ^{i-1} \left( \frac{\lambda_j}{\lambda_{j+1}} y (u) +1 \right) \prod_{k=i} ^{n-1} \left( y(u) + \frac{\lambda_k}{\lambda_{k+1}}\right) \lambda_n
\end{equation}
The two conditions mentioned for $f_l(u)$'s are still valid for $y(u)$'s. Now we can examine the Yang-Baxterization methods for different $n$ values. There are basically two main tools to do that: directly using the BGR's or using the relative algebras contained in them. Both cases will be discussed in this section. In both ways, the constraint on BGR's to be Yang-Baxterizable is to contain relevant $n$-CB algebra. Before moving on to a more detailed prescription for Yang-Baxterization in different $n$ values, the $n$-CB algebras will be introduced.

\subsection{$n$-CB algebras}
In the work of many authors \cite{cheng1991yang, li1993yang, belavin20194CB} it has been shown that $n$-block models satisfying the Yang-Baxter equation contain certain algebras, generally called $n$-conformal braiding ($n$-CB) algebras. As a consequence of this property, the inclusion of corresponding $n$-CB algebras is a sufficient condition for BGRs with $n$ different eigenvalues to be Yang-Baxterizable. There are known methods for Yang-Baxterizing a BGR containing these algebras, so it is interesting to study the generators and relations of these algebras.

Temperley and Lieb \cite{temperley2004relations} showed that the transfer matrices of seemingly different models can be written in terms of operators obeying the same algebra, the Temperley-Lieb (TL) algebra. After their paper, it was shown that every 2-block model obeys the TL algebra. Temperley and Lieb's work raised the question of what the most general algebras contained in any $n$-block model. In the recent work of Belavin and Gepner, these algebras have been studied for $n=3,4,5$. It was shown that $n$-CB algebras are obeyed by the Yang-Baxter operators if and only if the Yang-Baxter equation is obeyed by the $n$-block model, so these algebras inevitably play an important role in Yang-Baxterization. These $n$-block algebras form a chain of algebras such that each of them is a quotient of the universal free algebra generated by $\{1\}$, $\{E_i\}$, and $\{G_i\}$. The defining relations of the $n+1^{th}$ quotient contain the defining relations of the $n^{th}$ quotient except for the skein relation. Using the isomorphism between the general free algebra generated by $\{1\}$, $\{E_i\}$, and $\{G_i\}$ and Kauffman's tangle algebra \cite{kauffman1990invariant}, the relations of the $n$-CB algebras can be diagrammed. 

\subsubsection{$n=2$: Temperley-Lieb algebra}

Temperley Lieb algebra is the smallest non-trivial $n$-CB algebra. It is generated by $\{E_i\}$ such that $E_i$'s obey the relations:
\begin{align}
    &E_i E_j = E_j E_i; \ |i-j| \geq 2  \nonumber \\
    &E_i E_{i \pm 1} E_i = E_{i\pm 1}  \nonumber \\
    &E_i^2=\delta E_i
\end{align}
The generators and relations in this algebra can be illustrated using monograms. Their corresponding illustrations are given in Figure 11.
\tikzset{every picture/.style={line width=0.75pt}} 
\begin{center}
\begin{tikzpicture}[x=0.75pt,y=0.75pt,yscale=-1,xscale=1]

\draw  [draw opacity=0] (309.42,161.27) .. controls (310.09,151.05) and (318.56,142.95) .. (328.95,142.89) .. controls (339.43,142.83) and (348.04,150.95) .. (348.71,161.27) -- (329.07,162.58) -- cycle ; \draw   (309.42,161.27) .. controls (310.09,151.05) and (318.56,142.95) .. (328.95,142.89) .. controls (339.43,142.83) and (348.04,150.95) .. (348.71,161.27) ;  
\draw    (309.42,161.27) -- (309.42,172.82) ;
\draw    (348.71,161.27) -- (348.71,172.82) ;

\draw  [draw opacity=0] (348.69,117.59) .. controls (347.85,127.79) and (339.25,135.75) .. (328.85,135.64) .. controls (318.38,135.53) and (309.91,127.26) .. (309.4,116.93) -- (329.07,115.95) -- cycle ; \draw   (348.69,117.59) .. controls (347.85,127.79) and (339.25,135.75) .. (328.85,135.64) .. controls (318.38,135.53) and (309.91,127.26) .. (309.4,116.93) ;  
\draw    (348.69,117.59) -- (348.88,106.04) ;
\draw    (309.4,116.93) -- (309.59,105.39) ;

\draw (239.33,85.45) node [anchor=north west][inner sep=0.75pt]    {$1$};
\draw (264.64,85.45) node [anchor=north west][inner sep=0.75pt]    {$2$};
\draw (276.29,84.52) node [anchor=north west][inner sep=0.75pt]    {$\cdots $};
\draw (306.82,86.39) node [anchor=north west][inner sep=0.75pt]    {$i$};
\draw (332.74,85.49) node [anchor=north west][inner sep=0.75pt]    {$i+1$};
\draw (377.9,85.45) node [anchor=north west][inner sep=0.75pt]    {$\cdots $};
\draw (414.99,85.45) node [anchor=north west][inner sep=0.75pt]    {$n$};
\draw (238.3,171.08) node [anchor=north west][inner sep=0.75pt]    {$1'$};
\draw (263.61,171.08) node [anchor=north west][inner sep=0.75pt]    {$2'$};
\draw (276.29,170.15) node [anchor=north west][inner sep=0.75pt]    {$\cdots $};
\draw (306.82,172.02) node [anchor=north west][inner sep=0.75pt]    {$i'$};
\draw (331.69,172.02) node [anchor=north west][inner sep=0.75pt]    {$i+1'$};
\draw (377.9,171.08) node [anchor=north west][inner sep=0.75pt]    {$\cdots $};
\draw (413.93,171.08) node [anchor=north west][inner sep=0.75pt]    {$n'$};
\draw (208.86,124.77) node [anchor=north west][inner sep=0.75pt]    {$E_{i}$};
\draw    (245.39,105.05) -- (245.73,166.68) ;
\draw    (270.7,105.05) -- (271.04,166.68) ;
\draw    (421.12,105.05) -- (421.8,166.68) ;

\end{tikzpicture}

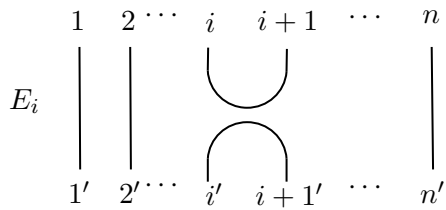
Figure 11: Diagrammatic representation of the generators of the TL algebra.
\end{center}
Using these monoid diagrams we can illustrate the defining relations of the TL algebra. These diagrams are shown in Figure 12.
\begin{center}

\tikzset{every picture/.style={line width=0.75pt}} 

\begin{tikzpicture}[x=0.75pt,y=0.75pt,yscale=-1,xscale=1]

\draw  [draw opacity=0] (45.5,180.92) .. controls (45.94,174.15) and (51.56,168.77) .. (58.45,168.73) .. controls (65.39,168.69) and (71.09,174.08) .. (71.54,180.92) -- (58.52,181.79) -- cycle ; \draw   (45.5,180.92) .. controls (45.94,174.15) and (51.56,168.77) .. (58.45,168.73) .. controls (65.39,168.69) and (71.09,174.08) .. (71.54,180.92) ;  
\draw    (45.5,180.92) -- (45.47,191.07) ;
\draw    (71.54,180.92) -- (71.57,190.63) ;

\draw  [draw opacity=0] (71.57,153.06) .. controls (71.17,159.83) and (65.59,165.24) .. (58.7,165.32) .. controls (51.75,165.4) and (46.01,160.05) .. (45.52,153.22) -- (58.54,152.27) -- cycle ; \draw   (71.57,153.06) .. controls (71.17,159.83) and (65.59,165.24) .. (58.7,165.32) .. controls (51.75,165.4) and (46.01,160.05) .. (45.52,153.22) ;  
\draw    (71.57,153.06) -- (71.32,108.55) ;
\draw    (45.52,153.22) -- (45.47,108.85) ;

\draw  [draw opacity=0] (102.7,118.31) .. controls (102.28,125.09) and (96.68,130.47) .. (89.79,130.53) .. controls (82.85,130.6) and (77.13,125.23) .. (76.66,118.39) -- (89.68,117.48) -- cycle ; \draw   (102.7,118.31) .. controls (102.28,125.09) and (96.68,130.47) .. (89.79,130.53) .. controls (82.85,130.6) and (77.13,125.23) .. (76.66,118.39) ;  
\draw    (102.7,118.31) -- (102.7,108.17) ;
\draw    (76.66,118.39) -- (76.6,108.68) ;

\draw  [draw opacity=0] (76.72,146.25) .. controls (77.1,139.47) and (82.66,134.05) .. (89.55,133.95) .. controls (96.49,133.85) and (102.25,139.18) .. (102.76,146.01) -- (89.75,147) -- cycle ; \draw   (76.72,146.25) .. controls (77.1,139.47) and (82.66,134.05) .. (89.55,133.95) .. controls (96.49,133.85) and (102.25,139.18) .. (102.76,146.01) ;  
\draw    (102.76,146.01) -- (102.89,191.07) ;
\draw    (76.72,146.25) -- (76.79,191.5) ;

\draw  [draw opacity=0] (159.91,181.35) .. controls (160.35,174.58) and (165.97,169.21) .. (172.86,169.17) .. controls (179.8,169.13) and (185.51,174.52) .. (185.95,181.35) -- (172.93,182.22) -- cycle ; \draw   (159.91,181.35) .. controls (160.35,174.58) and (165.97,169.21) .. (172.86,169.17) .. controls (179.8,169.13) and (185.51,174.52) .. (185.95,181.35) ;  
\draw    (159.91,181.35) -- (159.88,191.5) ;
\draw    (185.95,181.35) -- (185.98,191.07) ;

\draw  [draw opacity=0] (185.98,153.49) .. controls (185.58,160.27) and (180,165.68) .. (173.11,165.76) .. controls (166.17,165.84) and (160.43,160.49) .. (159.93,153.65) -- (172.95,152.71) -- cycle ; \draw   (185.98,153.49) .. controls (185.58,160.27) and (180,165.68) .. (173.11,165.76) .. controls (166.17,165.84) and (160.43,160.49) .. (159.93,153.65) ;  
\draw    (185.98,153.49) -- (185.74,108.98) ;
\draw    (159.93,153.65) -- (159.88,109.28) ;

\draw  [draw opacity=0] (154.9,118.31) .. controls (154.48,125.09) and (148.88,130.47) .. (141.99,130.53) .. controls (135.05,130.6) and (129.33,125.23) .. (128.86,118.39) -- (141.88,117.48) -- cycle ; \draw   (154.9,118.31) .. controls (154.48,125.09) and (148.88,130.47) .. (141.99,130.53) .. controls (135.05,130.6) and (129.33,125.23) .. (128.86,118.39) ;  
\draw    (154.9,118.31) -- (154.9,108.17) ;
\draw    (128.86,118.39) -- (128.8,108.68) ;

\draw  [draw opacity=0] (128.92,146.25) .. controls (129.3,139.47) and (134.86,134.05) .. (141.75,133.95) .. controls (148.7,133.85) and (154.45,139.18) .. (154.96,146.01) -- (141.95,147) -- cycle ; \draw   (128.92,146.25) .. controls (129.3,139.47) and (134.86,134.05) .. (141.75,133.95) .. controls (148.7,133.85) and (154.45,139.18) .. (154.96,146.01) ;  
\draw    (154.96,146.01) -- (155.1,191.07) ;
\draw    (128.92,146.25) -- (128.99,191.5) ;

\draw  [draw opacity=0] (351.56,162.99) .. controls (351.94,156.36) and (357.38,151.05) .. (364.13,150.95) .. controls (370.92,150.85) and (376.55,156.07) .. (377.05,162.76) -- (364.32,163.72) -- cycle ; \draw   (351.56,162.99) .. controls (351.94,156.36) and (357.38,151.05) .. (364.13,150.95) .. controls (370.92,150.85) and (376.55,156.07) .. (377.05,162.76) ;  
\draw    (377.05,162.76) -- (377.18,206.85) ;
\draw    (351.56,162.99) -- (351.64,207.28) ;

\draw  [draw opacity=0] (376.91,134.79) .. controls (376.48,141.42) and (370.98,146.67) .. (364.24,146.71) .. controls (357.44,146.75) and (351.86,141.47) .. (351.42,134.78) -- (364.17,133.94) -- cycle ; \draw   (376.91,134.79) .. controls (376.48,141.42) and (370.98,146.67) .. (364.24,146.71) .. controls (357.44,146.75) and (351.86,141.47) .. (351.42,134.78) ;  
\draw    (351.42,134.78) -- (351.72,90.69) ;
\draw    (376.91,134.79) -- (377.26,90.5) ;

\draw    (407.04,92.21) -- (406.24,207.4) ;
\draw  [draw opacity=0] (255.39,192.59) .. controls (255.92,185.97) and (261.49,180.79) .. (268.23,180.85) .. controls (275.03,180.91) and (280.54,186.26) .. (280.88,192.96) -- (268.12,193.62) -- cycle ; \draw   (255.39,192.59) .. controls (255.92,185.97) and (261.49,180.79) .. (268.23,180.85) .. controls (275.03,180.91) and (280.54,186.26) .. (280.88,192.96) ;  
\draw    (255.39,192.59) -- (255.3,208.2) ;
\draw    (280.88,192.96) -- (280.72,208.2) ;

\draw  [draw opacity=0] (255.39,133.01) .. controls (255.92,126.39) and (261.49,121.21) .. (268.23,121.27) .. controls (275.03,121.33) and (280.54,126.68) .. (280.88,133.37) -- (268.12,134.04) -- cycle ; \draw   (255.39,133.01) .. controls (255.92,126.39) and (261.49,121.21) .. (268.23,121.27) .. controls (275.03,121.33) and (280.54,126.68) .. (280.88,133.37) ;  
\draw    (255.39,133.01) -- (255.22,146.12) ;
\draw  [draw opacity=0] (280.87,165.8) .. controls (280.48,172.43) and (275.02,177.72) .. (268.27,177.8) .. controls (261.48,177.88) and (255.86,172.64) .. (255.38,165.95) -- (268.12,165.02) -- cycle ; \draw   (280.87,165.8) .. controls (280.48,172.43) and (275.02,177.72) .. (268.27,177.8) .. controls (261.48,177.88) and (255.86,172.64) .. (255.38,165.95) ;  
\draw    (255.38,165.95) -- (255.22,146.12) ;
\draw  [draw opacity=0] (280.87,105.67) .. controls (280.4,112.29) and (274.88,117.52) .. (268.13,117.52) .. controls (261.34,117.53) and (255.78,112.22) .. (255.38,105.53) -- (268.13,104.75) -- cycle ; \draw   (280.87,105.67) .. controls (280.4,112.29) and (274.88,117.52) .. (268.13,117.52) .. controls (261.34,117.53) and (255.78,112.22) .. (255.38,105.53) ;  
\draw    (280.87,105.67) -- (280.82,90.06) ;
\draw    (255.38,105.53) -- (255.4,90.29) ;

\draw  [draw opacity=0] (280.87,165.8) .. controls (281.4,159.18) and (286.97,154) .. (293.71,154.06) .. controls (300.51,154.11) and (306.02,159.47) .. (306.36,166.16) -- (293.6,166.83) -- cycle ; \draw   (280.87,165.8) .. controls (281.4,159.18) and (286.97,154) .. (293.71,154.06) .. controls (300.51,154.11) and (306.02,159.47) .. (306.36,166.16) ;  
\draw  [draw opacity=0] (306.36,133.88) .. controls (305.8,140.5) and (300.2,145.65) .. (293.46,145.55) .. controls (286.66,145.46) and (281.18,140.07) .. (280.88,133.37) -- (293.64,132.78) -- cycle ; \draw   (306.36,133.88) .. controls (305.8,140.5) and (300.2,145.65) .. (293.46,145.55) .. controls (286.66,145.46) and (281.18,140.07) .. (280.88,133.37) ;  
\draw    (306.94,91.42) -- (306.36,133.88) ;
\draw    (306.36,166.16) -- (306.14,207.4) ;

\draw  [draw opacity=0] (564.1,162.96) .. controls (564.57,155.69) and (570.6,149.93) .. (577.99,149.88) .. controls (585.44,149.84) and (591.57,155.62) .. (592.05,162.96) -- (578.07,163.89) -- cycle ; \draw   (564.1,162.96) .. controls (564.57,155.69) and (570.6,149.93) .. (577.99,149.88) .. controls (585.44,149.84) and (591.57,155.62) .. (592.05,162.96) ;  
\draw    (564.1,162.96) -- (564.09,171.18) ;
\draw    (592.05,162.96) -- (592.05,171.18) ;

\draw  [draw opacity=0] (592.03,131.89) .. controls (591.43,139.14) and (585.32,144.81) .. (577.92,144.73) .. controls (570.47,144.65) and (564.44,138.77) .. (564.08,131.42) -- (578.07,130.72) -- cycle ; \draw   (592.03,131.89) .. controls (591.43,139.14) and (585.32,144.81) .. (577.92,144.73) .. controls (570.47,144.65) and (564.44,138.77) .. (564.08,131.42) ;  
\draw    (592.03,131.89) -- (592.17,123.67) ;
\draw    (564.08,131.42) -- (564.22,123.21) ;

\draw  [draw opacity=0] (469.06,178.27) .. controls (469.54,171) and (475.56,165.23) .. (482.95,165.19) .. controls (490.4,165.15) and (496.53,170.93) .. (497.01,178.27) -- (483.04,179.2) -- cycle ; \draw   (469.06,178.27) .. controls (469.54,171) and (475.56,165.23) .. (482.95,165.19) .. controls (490.4,165.15) and (496.53,170.93) .. (497.01,178.27) ;  
\draw    (469.06,178.27) -- (469.06,186.49) ;
\draw    (497.01,178.27) -- (497.01,186.49) ;

\draw  [draw opacity=0] (496.99,118.49) .. controls (496.4,125.75) and (490.28,131.42) .. (482.88,131.34) .. controls (475.43,131.25) and (469.41,125.37) .. (469.05,118.03) -- (483.04,117.33) -- cycle ; \draw   (496.99,118.49) .. controls (496.4,125.75) and (490.28,131.42) .. (482.88,131.34) .. controls (475.43,131.25) and (469.41,125.37) .. (469.05,118.03) ;  
\draw    (496.99,118.49) -- (497.13,110.28) ;
\draw    (469.05,118.03) -- (469.18,109.81) ;

\draw   (469.06,147.95) .. controls (469.06,140.23) and (475.32,133.97) .. (483.04,133.97) .. controls (490.75,133.97) and (497.01,140.23) .. (497.01,147.95) .. controls (497.01,155.66) and (490.75,161.92) .. (483.04,161.92) .. controls (475.32,161.92) and (469.06,155.66) .. (469.06,147.95) -- cycle ;
\draw   (530.29,146.03) .. controls (530.29,138.31) and (536.55,132.06) .. (544.27,132.06) .. controls (551.98,132.06) and (558.24,138.31) .. (558.24,146.03) .. controls (558.24,153.75) and (551.98,160.01) .. (544.27,160.01) .. controls (536.55,160.01) and (530.29,153.75) .. (530.29,146.03) -- cycle ;

\draw (41.96,89.88) node [anchor=north west][inner sep=0.75pt]    {$i$};
\draw (73.29,88.76) node [anchor=north west][inner sep=0.75pt]    {$j$};
\draw (137.99,197.77) node [anchor=north west][inner sep=0.75pt]    {$E_{i} E_{j}$};
\draw (157.19,87.82) node [anchor=north west][inner sep=0.75pt]    {$j$};
\draw (104.5,139.3) node [anchor=north west][inner sep=0.75pt]    {$=$};
\draw (125.49,89.88) node [anchor=north west][inner sep=0.75pt]    {$i$};
\draw (54.9,197.34) node [anchor=north west][inner sep=0.75pt]    {$E_{j} E_{i}$};
\draw (347.81,74.02) node [anchor=north west][inner sep=0.75pt]    {$i$};
\draw (319.57,140.75) node [anchor=north west][inner sep=0.75pt]    {$=$};
\draw (369.62,214.18) node [anchor=north west][inner sep=0.75pt]    {$E_{i}$};
\draw (252.41,213.53) node [anchor=north west][inner sep=0.75pt]    {$E_{i} E_{i+1} E_{i}$};
\draw (251.68,73.22) node [anchor=north west][inner sep=0.75pt]    {$i$};
\draw (560.52,107.5) node [anchor=north west][inner sep=0.75pt]    {$i$};
\draw (465.48,94.11) node [anchor=north west][inner sep=0.75pt]    {$i$};
\draw (503.62,138.06) node [anchor=north west][inner sep=0.75pt]    {$=$};
\draw (474.29,192.61) node [anchor=north west][inner sep=0.75pt]    {$E_{i}^{2}$};
\draw (566.29,175.88) node [anchor=north west][inner sep=0.75pt]    {$\delta E_{i}$};

\end{tikzpicture}

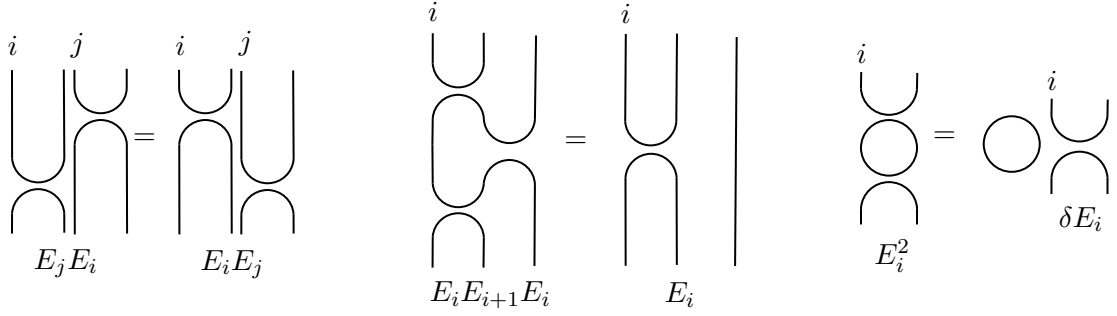
Figure 12: The diagrammatic representation of the defining relations of the TL algebra.
\end{center}

\subsubsection{$n=3$: BMW' algebra}

Birman and Wenlz, and Murakami independently showed in \cite{birman1989braids} and \cite{murakami1987kauffman} that some 3-block models obey an algebra known as Birman-Murakami-Wenlz (BMW) Algebra. The BMW algebra is a quotient algebra of the group algebra of braid groups \cite{rui2011blocks}. Yet the algebra was not generalized to be contained in every 3-block model. Later in 2019, Belavin and Gepner showed that an algebra called the "weak BMW Algebra" which is a subalgebra of BMW Algebra is contained in every 3-block model \cite{belavin20194CB}. This algebra is denoted by BMW' algebra and is generated by $\{G_i\}$, $\{E_i\}$, and $\{1_i\}$. The skein relation of this algebra differs from the skein relation of the BMW algebra and is the following.
\begin{equation}
    m(E_i-I) =G_i^{-1}-G_i
\end{equation}
$\{ G_i \}$'s satisfies the defining relations of the braid group, and $\{ E_i \}$'s satisfies the defining relations of the TL algebra. With the addition of the following relations between these two kinds of generators, the defining relations of the BMW' algebra are obtained. 
\begingroup
\allowdisplaybreaks
\begin{align}
&G_{i+1} G_i E_{i+1}= E_i G_{i+1} G_i=E_i E_{i+1} \nonumber \\
&G_{i+1}E_i G_{i+1}=G_i^{-1}E_{i+1}G_i^{-1} \nonumber \\
&G_{i+1}E_i E_{i+1}=G_i^{-1}E_{i+1} \nonumber \\
&G_i E_i = E_i G_i = l^{-1}E_i \nonumber \\
&E_i G_{i+1} E_i = l E_i
\end{align}
\endgroup
Remark that the BMW' algebra does not include $\{ G_i^{-1} \}$'s, however, for notational convenience, when describing the relations $\{ G_i^{-1} \}$'s are often used. \par
Using the relation between BMW' algebra and the tangle algebra, the relations of BMW' algebra are diagrammized and given in Figure 13.
\begin{center}

\tikzset{every picture/.style={line width=0.75pt}} 

\begin{tikzpicture}[x=0.75pt,y=0.75pt,yscale=-1,xscale=1]

\draw  [draw opacity=0] (230.69,143.49) .. controls (230.69,143.49) and (230.69,143.49) .. (230.69,143.49) .. controls (230.67,135.39) and (237.22,128.82) .. (245.32,128.81) .. controls (253.41,128.8) and (259.98,135.35) .. (260,143.44) .. controls (260,143.44) and (260,143.44) .. (260,143.44) -- (245.34,143.47) -- cycle ; \draw   (230.69,143.49) .. controls (230.69,143.49) and (230.69,143.49) .. (230.69,143.49) .. controls (230.67,135.39) and (237.22,128.82) .. (245.32,128.81) .. controls (253.41,128.8) and (259.98,135.35) .. (260,143.44) .. controls (260,143.44) and (260,143.44) .. (260,143.44) ;  
\draw  [draw opacity=0] (260,111.24) .. controls (260,111.24) and (260,111.24) .. (260,111.24) .. controls (260,111.24) and (260,111.24) .. (260,111.24) .. controls (259.89,119.34) and (253.24,125.81) .. (245.15,125.7) .. controls (237.06,125.6) and (230.58,118.95) .. (230.69,110.86) -- (245.34,111.05) -- cycle ; \draw   (260,111.24) .. controls (260,111.24) and (260,111.24) .. (260,111.24) .. controls (260,111.24) and (260,111.24) .. (260,111.24) .. controls (259.89,119.34) and (253.24,125.81) .. (245.15,125.7) .. controls (237.06,125.6) and (230.58,118.95) .. (230.69,110.86) ;  
\draw    (206.39,40.23) -- (206.39,72.64) ;
\draw    (259.65,74.38) -- (260,111.24) ;
\draw    (243.44,53.54) -- (259.65,74.38) ;
\draw    (243.44,60.1) -- (222.02,90.01) ;
\draw    (206.97,109.11) -- (216.81,96.38) ;
\draw    (232.44,38.49) -- (243.44,53.54) ;
\draw    (206.39,72.64) -- (230.69,110.86) ;
\draw    (206.97,109.11) -- (206.97,143.26) ;
\draw    (259.07,38.68) -- (248.65,53.73) ;

\draw  [draw opacity=0] (331.53,38.1) .. controls (331.53,38.1) and (331.53,38.1) .. (331.53,38.1) .. controls (331.55,46.19) and (325,52.76) .. (316.9,52.78) .. controls (308.81,52.79) and (302.24,46.24) .. (302.22,38.15) -- (316.88,38.12) -- cycle ; \draw   (331.53,38.1) .. controls (331.53,38.1) and (331.53,38.1) .. (331.53,38.1) .. controls (331.55,46.19) and (325,52.76) .. (316.9,52.78) .. controls (308.81,52.79) and (302.24,46.24) .. (302.22,38.15) ;  
\draw  [draw opacity=0] (302.23,70.35) .. controls (302.23,70.35) and (302.23,70.35) .. (302.23,70.35) .. controls (302.33,62.25) and (308.98,55.78) .. (317.07,55.88) .. controls (325.17,55.99) and (331.64,62.63) .. (331.54,70.73) -- (316.88,70.54) -- cycle ; \draw   (302.23,70.35) .. controls (302.23,70.35) and (302.23,70.35) .. (302.23,70.35) .. controls (302.33,62.25) and (308.98,55.78) .. (317.07,55.88) .. controls (325.17,55.99) and (331.64,62.63) .. (331.54,70.73) ;  
\draw    (355.84,141.35) -- (355.84,108.94) ;
\draw    (302.58,107.21) -- (302.23,70.35) ;
\draw    (318.8,128.05) -- (302.58,107.21) ;
\draw    (318.79,121.49) -- (340.21,91.58) ;
\draw    (355.25,72.47) -- (345.41,85.21) ;
\draw    (329.8,143.09) -- (318.8,128.05) ;
\draw    (355.84,108.94) -- (331.54,70.73) ;
\draw    (355.25,72.47) -- (355.25,38.32) ;
\draw    (303.17,142.91) -- (313.59,127.85) ;
\draw  [draw opacity=0] (425.63,38.1) .. controls (425.63,38.1) and (425.63,38.1) .. (425.63,38.1) .. controls (425.65,46.19) and (419.1,52.76) .. (411,52.78) .. controls (402.91,52.79) and (396.33,46.24) .. (396.32,38.15) -- (410.98,38.12) -- cycle ; \draw   (425.63,38.1) .. controls (425.63,38.1) and (425.63,38.1) .. (425.63,38.1) .. controls (425.65,46.19) and (419.1,52.76) .. (411,52.78) .. controls (402.91,52.79) and (396.33,46.24) .. (396.32,38.15) ;  
\draw  [draw opacity=0] (396.32,71.15) .. controls (396.32,71.15) and (396.32,71.15) .. (396.32,71.15) .. controls (396.42,63.49) and (402.7,57.37) .. (410.35,57.47) .. controls (418,57.57) and (424.13,63.85) .. (424.03,71.5) -- (410.17,71.33) -- cycle ; \draw   (396.32,71.15) .. controls (396.32,71.15) and (396.32,71.15) .. (396.32,71.15) .. controls (396.42,63.49) and (402.7,57.37) .. (410.35,57.47) .. controls (418,57.57) and (424.13,63.85) .. (424.03,71.5) ;  
\draw    (396.68,107.21) -- (396.33,70.35) ;
\draw    (424.47,107.84) -- (424.03,71.5) ;
\draw    (449.93,73.22) -- (449.93,38.49) ;
\draw  [draw opacity=0] (423.86,142.33) .. controls (423.86,142.33) and (423.86,142.33) .. (423.86,142.33) .. controls (423.85,134.24) and (430.4,127.67) .. (438.5,127.65) .. controls (446.59,127.64) and (453.16,134.19) .. (453.18,142.28) -- (438.52,142.31) -- cycle ; \draw   (423.86,142.33) .. controls (423.86,142.33) and (423.86,142.33) .. (423.86,142.33) .. controls (423.85,134.24) and (430.4,127.67) .. (438.5,127.65) .. controls (446.59,127.64) and (453.16,134.19) .. (453.18,142.28) ;  
\draw  [draw opacity=0] (450.07,108.18) .. controls (450.07,108.18) and (450.07,108.18) .. (450.07,108.18) .. controls (450.07,108.18) and (450.07,108.18) .. (450.07,108.18) .. controls (449.97,115.25) and (444.17,120.9) .. (437.1,120.81) .. controls (430.03,120.72) and (424.38,114.91) .. (424.47,107.84) -- (437.27,108.01) -- cycle ; \draw   (450.07,108.18) .. controls (450.07,108.18) and (450.07,108.18) .. (450.07,108.18) .. controls (450.07,108.18) and (450.07,108.18) .. (450.07,108.18) .. controls (449.97,115.25) and (444.17,120.9) .. (437.1,120.81) .. controls (430.03,120.72) and (424.38,114.91) .. (424.47,107.84) ;  
\draw    (449.93,73.22) -- (450.28,110.08) ;
\draw    (396.68,107.21) -- (396.1,139.79) ;

\draw (204.45,94.95) node [anchor=north west][inner sep=0.75pt]    {$\ $};
\draw (440.98,63.11) node [anchor=north west][inner sep=0.75pt]    {$\ \ \ $};
\draw (268.15,81.08) node [anchor=north west][inner sep=0.75pt]    {$=$};
\draw (364.1,81.86) node [anchor=north west][inner sep=0.75pt]    {$=$};
\draw (400.05,154.24) node [anchor=north west][inner sep=0.75pt]    {$E_i E_{i+1}$};
\draw (296.68,154.24) node [anchor=north west][inner sep=0.75pt]    {$E_i G_{i+1} G_i$};
\draw (198.84,154.24) node [anchor=north west][inner sep=0.75pt]    {$G_{i+1} G_i E_{i+1}$};
\draw (311.53,117.32) node [anchor=north west][inner sep=0.75pt]  [rotate=-179.99]  {$\ \ \ $};
\draw (250.7,64.27) node [anchor=north west][inner sep=0.75pt]    {$\ \ \ $};

\end{tikzpicture}

\hfill \break

\tikzset{every picture/.style={line width=0.75pt}} 

\begin{tikzpicture}[x=0.75pt,y=0.75pt,yscale=-1,xscale=1]

\draw  [draw opacity=0] (101.46,98.57) .. controls (101.46,98.57) and (101.46,98.57) .. (101.46,98.57) .. controls (101.46,98.57) and (101.46,98.57) .. (101.46,98.57) .. controls (101.45,90.48) and (108,83.9) .. (116.09,83.89) .. controls (124.19,83.88) and (130.76,90.43) .. (130.77,98.52) -- (116.12,98.55) -- cycle ; \draw   (101.46,98.57) .. controls (101.46,98.57) and (101.46,98.57) .. (101.46,98.57) .. controls (101.46,98.57) and (101.46,98.57) .. (101.46,98.57) .. controls (101.45,90.48) and (108,83.9) .. (116.09,83.89) .. controls (124.19,83.88) and (130.76,90.43) .. (130.77,98.52) ;  
\draw  [draw opacity=0] (130.83,63.65) .. controls (130.83,63.65) and (130.83,63.65) .. (130.83,63.65) .. controls (130.83,63.65) and (130.83,63.65) .. (130.83,63.65) .. controls (130.73,71.74) and (124.08,78.22) .. (115.99,78.11) .. controls (107.89,78.01) and (101.42,71.36) .. (101.52,63.27) -- (116.18,63.46) -- cycle ; \draw   (130.83,63.65) .. controls (130.83,63.65) and (130.83,63.65) .. (130.83,63.65) .. controls (130.83,63.65) and (130.83,63.65) .. (130.83,63.65) .. controls (130.73,71.74) and (124.08,78.22) .. (115.99,78.11) .. controls (107.89,78.01) and (101.42,71.36) .. (101.52,63.27) ;  
\draw    (101.39,29.54) -- (101.52,67.44) ;
\draw    (154.64,63.7) -- (154.79,93.52) ;
\draw    (138.43,42.86) -- (154.64,63.7) ;
\draw    (139.76,50.1) -- (130.83,63.65) ;
\draw    (145.61,107.71) -- (154.79,93.52) ;
\draw    (127.43,27.81) -- (138.43,42.86) ;
\draw    (130.77,98.52) -- (155.63,130.26) ;
\draw    (101.46,98.57) -- (101.35,133.6) ;
\draw    (155.63,28.39) -- (144.48,44.72) ;
\draw    (125.57,131.93) -- (137.26,116.9) ;
\draw  [draw opacity=0] (249.6,61.32) .. controls (249.6,61.32) and (249.6,61.32) .. (249.6,61.32) .. controls (249.6,61.32) and (249.6,61.32) .. (249.6,61.32) .. controls (249.57,69.41) and (242.98,75.95) .. (234.89,75.92) .. controls (226.8,75.89) and (220.26,69.31) .. (220.29,61.21) -- (234.94,61.27) -- cycle ; \draw   (249.6,61.32) .. controls (249.6,61.32) and (249.6,61.32) .. (249.6,61.32) .. controls (249.6,61.32) and (249.6,61.32) .. (249.6,61.32) .. controls (249.57,69.41) and (242.98,75.95) .. (234.89,75.92) .. controls (226.8,75.89) and (220.26,69.31) .. (220.29,61.21) ;  
\draw  [draw opacity=0] (220.04,96.08) .. controls (220.04,96.08) and (220.04,96.08) .. (220.04,96.08) .. controls (220.04,96.08) and (220.04,96.08) .. (220.04,96.08) .. controls (220.19,87.99) and (226.87,81.55) .. (234.97,81.7) .. controls (243.06,81.85) and (249.5,88.53) .. (249.35,96.62) -- (234.7,96.35) -- cycle ; \draw   (220.04,96.08) .. controls (220.04,96.08) and (220.04,96.08) .. (220.04,96.08) .. controls (220.04,96.08) and (220.04,96.08) .. (220.04,96.08) .. controls (220.19,87.99) and (226.87,81.55) .. (234.97,81.7) .. controls (243.06,81.85) and (249.5,88.53) .. (249.35,96.62) ;  
\draw    (249.31,130.35) -- (249.37,92.45) ;
\draw    (196.24,95.92) -- (196.24,66.09) ;
\draw    (206.56,110.22) -- (196.24,95.92) ;
\draw    (213.24,50.93) -- (220.29,61.21) ;
\draw    (224.1,131.09) -- (211.57,116.06) ;
\draw    (196.24,66.09) -- (225.67,27.84) ;
\draw    (249.6,61.32) -- (249.89,26.29) ;
\draw    (195.07,131.22) -- (220.04,96.08) ;
\draw    (207.4,44.25) -- (194.87,26.72) ;
\draw  [draw opacity=0] (411.57,96.51) .. controls (411.57,96.51) and (411.57,96.51) .. (411.57,96.51) .. controls (411.57,96.51) and (411.57,96.51) .. (411.57,96.51) .. controls (411.56,88.65) and (417.91,82.28) .. (425.77,82.26) .. controls (433.62,82.25) and (440,88.61) .. (440.01,96.46) -- (425.79,96.49) -- cycle ; \draw   (411.57,96.51) .. controls (411.57,96.51) and (411.57,96.51) .. (411.57,96.51) .. controls (411.57,96.51) and (411.57,96.51) .. (411.57,96.51) .. controls (411.56,88.65) and (417.91,82.28) .. (425.77,82.26) .. controls (433.62,82.25) and (440,88.61) .. (440.01,96.46) ;  
\draw  [draw opacity=0] (440.94,62.02) .. controls (440.84,70.12) and (434.19,76.59) .. (426.1,76.48) .. controls (418,76.38) and (411.53,69.73) .. (411.63,61.64) -- (426.29,61.83) -- cycle ; \draw   (440.94,62.02) .. controls (440.84,70.12) and (434.19,76.59) .. (426.1,76.48) .. controls (418,76.38) and (411.53,69.73) .. (411.63,61.64) ;  
\draw    (411.49,27.92) -- (411.63,65.81) ;
\draw    (464.75,62.07) -- (465.62,102.55) ;
\draw    (448.54,41.23) -- (464.75,62.07) ;
\draw    (449.87,48.47) -- (440.94,62.02) ;
\draw    (437.54,26.18) -- (448.54,41.23) ;
\draw    (411.57,96.94) -- (411.46,131.97) ;
\draw    (465.73,26.76) -- (454.58,43.09) ;
\draw    (440.02,102.21) -- (440.01,96.46) ;
\draw  [draw opacity=0] (439.42,133.36) .. controls (439.42,133.36) and (439.42,133.36) .. (439.42,133.36) .. controls (439.41,125.26) and (445.96,118.69) .. (454.05,118.68) .. controls (462.15,118.67) and (468.72,125.22) .. (468.73,133.31) -- (454.07,133.34) -- cycle ; \draw   (439.42,133.36) .. controls (439.42,133.36) and (439.42,133.36) .. (439.42,133.36) .. controls (439.41,125.26) and (445.96,118.69) .. (454.05,118.68) .. controls (462.15,118.67) and (468.72,125.22) .. (468.73,133.31) ;  
\draw  [draw opacity=0] (465.62,102.55) .. controls (465.62,102.55) and (465.62,102.55) .. (465.62,102.55) .. controls (465.53,109.62) and (459.72,115.27) .. (452.65,115.18) .. controls (445.59,115.09) and (439.93,109.28) .. (440.02,102.21) -- (452.82,102.38) -- cycle ; \draw   (465.62,102.55) .. controls (465.62,102.55) and (465.62,102.55) .. (465.62,102.55) .. controls (465.53,109.62) and (459.72,115.27) .. (452.65,115.18) .. controls (445.59,115.09) and (439.93,109.28) .. (440.02,102.21) ;  
\draw  [draw opacity=0] (559.71,59.69) .. controls (559.68,67.79) and (553.09,74.32) .. (545,74.29) .. controls (536.9,74.26) and (530.37,67.68) .. (530.4,59.58) -- (545.05,59.64) -- cycle ; \draw   (559.71,59.69) .. controls (559.68,67.79) and (553.09,74.32) .. (545,74.29) .. controls (536.9,74.26) and (530.37,67.68) .. (530.4,59.58) ;  
\draw  [draw opacity=0] (530.15,94.46) .. controls (530.3,86.36) and (536.98,79.92) .. (545.07,80.07) .. controls (553.17,80.22) and (559.61,86.9) .. (559.46,94.99) -- (544.81,94.72) -- cycle ; \draw   (530.15,94.46) .. controls (530.3,86.36) and (536.98,79.92) .. (545.07,80.07) .. controls (553.17,80.22) and (559.61,86.9) .. (559.46,94.99) ;  
\draw    (559.26,130.86) -- (559.48,90.82) ;
\draw    (506.35,94.29) -- (506.35,64.46) ;
\draw    (530.03,130.02) -- (530.15,94.46) ;
\draw    (523.35,49.3) -- (530.4,59.58) ;
\draw    (506.35,64.46) -- (535.78,26.21) ;
\draw    (559.71,59.69) -- (560,24.66) ;
\draw    (506.65,130.02) -- (506.35,94.29) ;
\draw    (517.51,42.62) -- (504.98,25.09) ;

\draw (163.14,70.4) node [anchor=north west][inner sep=0.75pt]    {$=$};
\draw (90.33,141.72) node [anchor=north west][inner sep=0.75pt]    {$G_{i+1}E_i G_{i+1}$};
\draw (145.69,53.59) node [anchor=north west][inner sep=0.75pt]    {$\ \ \ $};
\draw (185.18,142.55) node [anchor=north west][inner sep=0.75pt]    {$G_i^{-1}E_{i+1}G_i^{-1}$};
\draw (473.25,68.77) node [anchor=north west][inner sep=0.75pt]    {$=$};
\draw (400.44,140.09) node [anchor=north west][inner sep=0.75pt]    {$G_{i+1}E_i E_{i+1}$};
\draw (455.8,51.96) node [anchor=north west][inner sep=0.75pt]    {$\ \ \ $};
\draw (495.79,140.93) node [anchor=north west][inner sep=0.75pt]    {$G_i^{-1}E_{i+1}$};

\end{tikzpicture}

\tikzset{every picture/.style={line width=0.75pt}} 

\begin{tikzpicture}[x=0.75pt,y=0.75pt,yscale=-1,xscale=1]

\draw    (104.24,74.94) -- (120.45,95.78) ;
\draw    (102.23,79.95) -- (91.14,95.45) ;
\draw    (91,57.55) -- (104.24,74.94) ;
\draw    (120.31,57.88) -- (108.91,73.27) ;
\draw  [draw opacity=0] (90.94,128.66) .. controls (90.94,128.66) and (90.94,128.66) .. (90.94,128.66) .. controls (90.93,120.56) and (97.48,113.99) .. (105.58,113.98) .. controls (113.67,113.97) and (120.24,120.52) .. (120.26,128.61) -- (105.6,128.64) -- cycle ; \draw   (90.94,128.66) .. controls (90.94,128.66) and (90.94,128.66) .. (90.94,128.66) .. controls (90.93,120.56) and (97.48,113.99) .. (105.58,113.98) .. controls (113.67,113.97) and (120.24,120.52) .. (120.26,128.61) ;  
\draw  [draw opacity=0] (120.45,95.78) .. controls (120.45,95.78) and (120.45,95.78) .. (120.45,95.78) .. controls (120.36,103.87) and (113.72,110.36) .. (105.63,110.27) .. controls (97.54,110.18) and (91.05,103.54) .. (91.14,95.45) -- (105.79,95.61) -- cycle ; \draw   (120.45,95.78) .. controls (120.45,95.78) and (120.45,95.78) .. (120.45,95.78) .. controls (120.36,103.87) and (113.72,110.36) .. (105.63,110.27) .. controls (97.54,110.18) and (91.05,103.54) .. (91.14,95.45) ;  
\draw    (172.43,111.82) -- (156.61,90.69) ;
\draw    (174.53,106.85) -- (185.9,91.55) ;
\draw    (185.35,129.45) -- (172.43,111.82) ;
\draw    (156.05,128.58) -- (167.73,113.4) ;
\draw  [draw opacity=0] (186.71,58.35) .. controls (186.71,58.35) and (186.71,58.35) .. (186.71,58.35) .. controls (186.57,66.45) and (179.9,72.9) .. (171.81,72.76) .. controls (163.71,72.63) and (157.26,65.96) .. (157.4,57.86) -- (172.05,58.11) -- cycle ; \draw   (186.71,58.35) .. controls (186.71,58.35) and (186.71,58.35) .. (186.71,58.35) .. controls (186.57,66.45) and (179.9,72.9) .. (171.81,72.76) .. controls (163.71,72.63) and (157.26,65.96) .. (157.4,57.86) ;  
\draw  [draw opacity=0] (156.61,90.69) .. controls (156.61,90.69) and (156.61,90.69) .. (156.61,90.69) .. controls (156.84,82.6) and (163.6,76.23) .. (171.69,76.47) .. controls (179.78,76.71) and (186.14,83.46) .. (185.9,91.55) .. controls (185.9,91.55) and (185.9,91.55) .. (185.9,91.55) -- (171.25,91.12) -- cycle ; \draw   (156.61,90.69) .. controls (156.61,90.69) and (156.61,90.69) .. (156.61,90.69) .. controls (156.84,82.6) and (163.6,76.23) .. (171.69,76.47) .. controls (179.78,76.71) and (186.14,83.46) .. (185.9,91.55) .. controls (185.9,91.55) and (185.9,91.55) .. (185.9,91.55) ;  
\draw  [draw opacity=0] (249.91,74.94) .. controls (254.21,72.19) and (259.83,71.77) .. (264.66,74.33) .. controls (271.82,78.11) and (274.55,86.98) .. (270.77,94.13) .. controls (266.99,101.29) and (258.12,104.02) .. (250.97,100.24) .. controls (246.17,97.7) and (243.36,92.88) .. (243.17,87.82) -- (257.81,87.28) -- cycle ; \draw   (249.91,74.94) .. controls (254.21,72.19) and (259.83,71.77) .. (264.66,74.33) .. controls (271.82,78.11) and (274.55,86.98) .. (270.77,94.13) .. controls (266.99,101.29) and (258.12,104.02) .. (250.97,100.24) .. controls (246.17,97.7) and (243.36,92.88) .. (243.17,87.82) ;  
\draw  [draw opacity=0] (249.91,74.94) .. controls (245.38,79.67) and (238.07,80.93) .. (232.11,77.56) .. controls (225.87,74.03) and (223.21,66.61) .. (225.45,60.06) -- (239.32,64.8) -- cycle ; \draw   (249.91,74.94) .. controls (245.38,79.67) and (238.07,80.93) .. (232.11,77.56) .. controls (225.87,74.03) and (223.21,66.61) .. (225.45,60.06) ;  
\draw    (244.18,72.16) -- (245.02,60.47) ;
\draw  [draw opacity=0] (222.88,128.66) .. controls (222.88,128.66) and (222.88,128.66) .. (222.88,128.66) .. controls (222.86,120.56) and (229.41,113.99) .. (237.51,113.98) .. controls (245.6,113.97) and (252.17,120.52) .. (252.19,128.61) -- (237.53,128.64) -- cycle ; \draw   (222.88,128.66) .. controls (222.88,128.66) and (222.88,128.66) .. (222.88,128.66) .. controls (222.86,120.56) and (229.41,113.99) .. (237.51,113.98) .. controls (245.6,113.97) and (252.17,120.52) .. (252.19,128.61) ;  
\draw    (415.73,110.74) -- (415.74,74.61) ;
\draw  [draw opacity=0] (445.04,41.59) .. controls (445.04,41.59) and (445.04,41.59) .. (445.04,41.59) .. controls (445,49.69) and (438.41,56.21) .. (430.31,56.17) .. controls (422.22,56.13) and (415.69,49.53) .. (415.73,41.44) -- (430.39,41.52) -- cycle ; \draw   (445.04,41.59) .. controls (445.04,41.59) and (445.04,41.59) .. (445.04,41.59) .. controls (445,49.69) and (438.41,56.21) .. (430.31,56.17) .. controls (422.22,56.13) and (415.69,49.53) .. (415.73,41.44) ;  
\draw  [draw opacity=0] (415.74,74.61) .. controls (415.61,66.51) and (422.07,59.85) .. (430.17,59.73) .. controls (438.26,59.6) and (444.92,66.06) .. (445.04,74.16) -- (430.39,74.38) -- cycle ; \draw   (415.74,74.61) .. controls (415.61,66.51) and (422.07,59.85) .. (430.17,59.73) .. controls (438.26,59.6) and (444.92,66.06) .. (445.04,74.16) ;  
\draw  [draw opacity=0] (445.04,110.9) .. controls (445.04,110.9) and (445.04,110.9) .. (445.04,110.9) .. controls (445,118.99) and (438.41,125.52) .. (430.31,125.48) .. controls (422.22,125.43) and (415.69,118.84) .. (415.73,110.74) -- (430.39,110.82) -- cycle ; \draw   (445.04,110.9) .. controls (445.04,110.9) and (445.04,110.9) .. (445.04,110.9) .. controls (445,118.99) and (438.41,125.52) .. (430.31,125.48) .. controls (422.22,125.43) and (415.69,118.84) .. (415.73,110.74) ;  
\draw  [draw opacity=0] (415.74,143.91) .. controls (415.61,135.82) and (422.07,129.16) .. (430.17,129.03) .. controls (438.26,128.91) and (444.92,135.37) .. (445.04,143.46) -- (430.39,143.69) -- cycle ; \draw   (415.74,143.91) .. controls (415.61,135.82) and (422.07,129.16) .. (430.17,129.03) .. controls (438.26,128.91) and (444.92,135.37) .. (445.04,143.46) ;  
\draw    (455.28,94.7) -- (445.04,110.9) ;
\draw    (445.04,74.16) -- (470.31,108.06) ;
\draw    (470.85,71.51) -- (460.29,87.18) ;
\draw    (470.85,71.51) -- (471.14,42.09) ;
\draw    (470.31,143.13) -- (470.31,108.06) ;
\draw  [draw opacity=0] (560.7,96.11) .. controls (559.4,100.91) and (555.68,104.94) .. (550.54,106.39) .. controls (542.76,108.6) and (534.65,104.08) .. (532.45,96.29) .. controls (530.24,88.51) and (534.76,80.4) .. (542.55,78.19) .. controls (547.77,76.71) and (553.14,78.26) .. (556.77,81.79) -- (546.55,92.29) -- cycle ; \draw   (560.7,96.11) .. controls (559.4,100.91) and (555.68,104.94) .. (550.54,106.39) .. controls (542.76,108.6) and (534.65,104.08) .. (532.45,96.29) .. controls (530.24,88.51) and (534.76,80.4) .. (542.55,78.19) .. controls (547.77,76.71) and (553.14,78.26) .. (556.77,81.79) ;  
\draw  [draw opacity=0] (556.77,81.79) .. controls (560.66,85.43) and (564.17,92.85) .. (565.79,101.77) .. controls (566.44,105.39) and (566.72,108.88) .. (566.66,112.08) -- (555.23,103.69) -- cycle ; \draw   (556.77,81.79) .. controls (560.66,85.43) and (564.17,92.85) .. (565.79,101.77) .. controls (566.44,105.39) and (566.72,108.88) .. (566.66,112.08) ;  
\draw    (564.66,86.62) -- (565.51,74.93) ;
\draw    (565.51,74.93) -- (565.5,44.04) ;
\draw    (566.37,141.5) -- (566.66,112.08) ;
\draw  [draw opacity=0] (508.28,118.91) .. controls (508.27,110.82) and (514.82,104.24) .. (522.92,104.23) .. controls (531.01,104.22) and (537.58,110.77) .. (537.6,118.86) -- (522.94,118.89) -- cycle ; \draw   (508.28,118.91) .. controls (508.27,110.82) and (514.82,104.24) .. (522.92,104.23) .. controls (531.01,104.22) and (537.58,110.77) .. (537.6,118.86) ;  
\draw  [draw opacity=0] (537.79,65.99) .. controls (537.79,65.99) and (537.79,65.99) .. (537.79,65.99) .. controls (537.7,74.08) and (531.06,80.57) .. (522.97,80.48) .. controls (514.88,80.39) and (508.39,73.76) .. (508.48,65.66) -- (523.13,65.83) -- cycle ; \draw   (537.79,65.99) .. controls (537.79,65.99) and (537.79,65.99) .. (537.79,65.99) .. controls (537.7,74.08) and (531.06,80.57) .. (522.97,80.48) .. controls (514.88,80.39) and (508.39,73.76) .. (508.48,65.66) ;  
\draw    (507.88,140.9) -- (508.28,118.91) ;
\draw    (537.94,141.74) -- (537.6,118.86) ;
\draw    (508.48,65.66) -- (508.88,43.67) ;
\draw    (537.79,65.99) -- (538.19,44) ;

\draw (165.37,100.96) node [anchor=north west][inner sep=0.75pt]  [rotate=-181.05]  {$\ \ \ $};
\draw (126.45,84.94) node [anchor=north west][inner sep=0.75pt]    {$=$};
\draw (196.59,84.94) node [anchor=north west][inner sep=0.75pt]    {$=$};
\draw (87.32,140.4) node [anchor=north west][inner sep=0.75pt]    {$G_i E_i$};
\draw (153.29,140.4) node [anchor=north west][inner sep=0.75pt]    {$E_i G_i$};
\draw (220.09,140.4) node [anchor=north west][inner sep=0.75pt]    {$l^{-1}E_i$};
\draw (477.82,83.55) node [anchor=north west][inner sep=0.75pt]    {$=$};
\draw (412.17,151.37) node [anchor=north west][inner sep=0.75pt]    {$E_i G_{i+1} E_i$};
\draw (526.36,152.21) node [anchor=north west][inner sep=0.75pt]    {$l E_i$};

\end{tikzpicture}

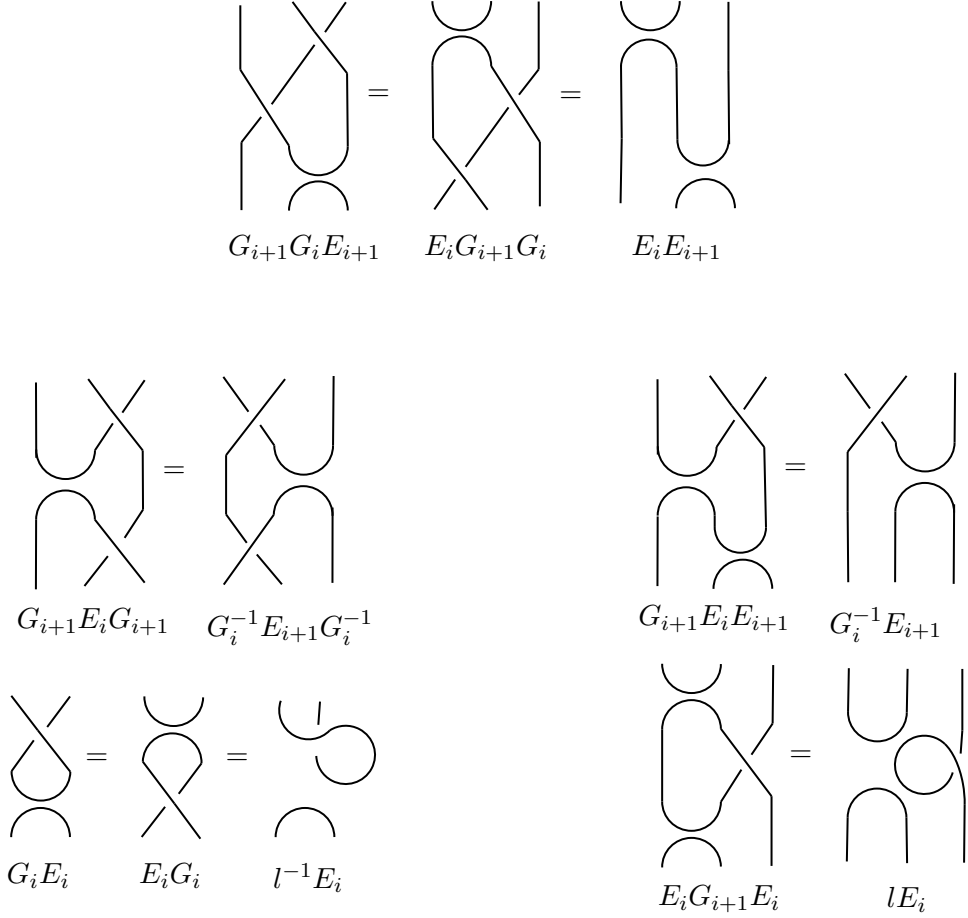
Figure 13: The diagrammatic representation of the defining relations of the BMW' algebra.
\end{center}
\subsubsection{$n=4$: 4-CB algebra}

For 4-block models, the corresponding algebra is the 4-CB algebra \cite{belavin20194CB}. The generators of 4-CB algebra are $\{G_i\}$, $\{G_i^{-1}\}$, $\{E_i\}$, and $\{1_i\}$. The skein relation of this algebra is the following.
\begin{equation}
    G_i^2=\alpha + \beta E_i + \gamma G_i + \delta G_i^{-1}
\end{equation}
where the coefficients $\alpha$, $\beta$, $\gamma$, and $\delta$ are parameters of the algebra.\par
There are 37 relations between the generators of this algebra. 4-CB contains both BMW' and TL algebras as subalgebras, hence some of these relations are the relations of BMW' and TL algebra. All of the relations can be shown in diagrams by using the relation between the tangle algebra and the generators of $4$-CB algebras, these diagrams are given in the appendix. Additionally, the additional relations for the $5$-CB algebras are also diagrammized and given in the appendix.

\subsection{Yang-Baxterization process of BGR's}

As mentioned in the previous sections, BGR's can be categorized with respect to the number of eigenvalues a $B$ matrix of a representation of the braid group has. This classification is often useful in the transition between Yang-Baxter operators and BGR's. Using the generators of the corresponding algebra, receipts for Yang-Baxterization can be written for each $n$ value. In this section, $n=2,3$ cases will be discussed and an example of the $n=2$ case will be given.

\subsection{$n=2$ case} 

Let us assume that the starting BGR has two distinct eigenvalues, $\lambda_1$ and $\lambda_2$. Using () and again assuming $f_l(u)=y(u)$, the Yang-Baxter operator has the form
\begin{equation}
    \check{R}(u) = B+\lambda_1 \lambda_2 y(u)B^{-1}
\end{equation}
In this case, BGR is directly used, another way to Yang-Baxterize BGR in $n=2$ case is to use the Temperley-Lieb algebra contained in it \cite{li1993yang}. $B$ can written as
\begin{align}
    B &= \lambda_1 P_1 + \lambda_2 P_2 \nonumber \\
    &=\lambda_1 (P_1 + \frac{\lambda_2}{\lambda_1} P_2) \nonumber \\
    &= \lambda_1 (I+ (\frac{\lambda_2}{\lambda_1} - 1 )P_2)
\end{align}
Omitting the overall constant $\lambda_1$, $B$ can be written as $B=I+\xi E$ where $E=P_2$ and $\xi = \frac{\lambda_2}{\lambda_1} - 1 $.
Substituting $B=I+\xi E$ in the YBE:
\begingroup
\allowdisplaybreaks
\begin{align}
& (I+\xi E_i)(I+\xi E_{i+1})(I+\xi E_i)=(I+\xi E_{i+1})(I+\xi E_i)(I+\xi E_{i+1}) \nonumber \\ 
=&(I+2 \xi E_i+\xi E_{i+1}+\xi^2 E_i^2 + \xi^2 E_i E_{i+1}+\xi ^2 E_{i+1} E_i +  \xi^3 E_i E_{i+1} E_i ) \nonumber \\
=&(I+ \xi E_i+ 2\xi E_{i+1}+\xi^2 E_{i+1}^2 + \xi^2 E_i E_{i+1}+\xi ^2 E_{i+1} E_i +  \xi^3E_{i+1}E_i E_{i+1} ) 
\end{align}
\endgroup
Therefore,
\begin{equation}
    \xi E_i+ \xi^2 E_i^2 +  \xi^3 E_i E_{i+1} E_i = \xi E_{i+1}+\xi^2 E_{i+1}^2 + \xi^3E_{i+1}E_i E_{i+1}
\end{equation}
It can be seen that if the Temperley-Lieb algebra is obeyed, then we can obtain
\begin{align}
    &\xi E_i+ \xi^2 \delta E_i +  \xi^3 E_i  = \xi E_{i+1}+\xi^2 
    \delta E_{i+1} + \xi^3 E_{i+1} \nonumber \\
    &(\xi + \xi^2 \delta + \xi^3 )E_i  = (\xi +\xi^2 
    \delta  + \xi^3 )E_{i+1}
\end{align}
With an appropriate choice of spectral function, the generators of the TL algebra can be used to derive solutions of the YBE. We now have a new tool in hand for Yang-Baxterization. The Yang-Baxter operator can be written as 
\begin{equation}
    \check{R}(u)=F(u)(I+ G(u)\xi E)
\end{equation}
It has been demonstrated in \cite{akutsu1989yang} that the Yang-Baxter operators obey the Temperley-Lieb algebra  at $u=\lambda$ (where $\lambda$ is the crossing parameter) in the $n \geq 2$ block cases. Therefore containing TL algebra serves as a sufficient condition for a BGR with two unequal eigenvalues to be Yang-Baxterizable \cite{cheng1991yang}.

\subsection{$n=3$ case}

Now suppose that the starting BGR has three distinct eigenvalues, $\lambda_1$, $\lambda_2$, and $\lambda_3$. Then $\check{R}(u)$ can written as 
\begin{equation}
    \check{R}(u) = L(u)B+M(u)I+N(u)B^{-1}
\end{equation}
where
\begin{align*}
L(u)=& 1-y(u)\\
M(u)=& (\lambda_1+\lambda_2)(\lambda_2+\lambda_3)\lambda_2^{-1}y(u)\\
N(u)=& \lambda_1 \lambda_3 y(u) [y(u)-1]
\end{align*}
Another approach is to use the relevant algebra for the $n=3$ case. It is known that $B$ satisfies the following equation
\begin{equation} \label{b2}
    B^2 =(\lambda_1+\lambda_2+\lambda_3) B-(\lambda_1 \lambda_2+\lambda_1 \lambda_3+\lambda_2 \lambda_3)I+\lambda_1 \lambda_2 \lambda_3 B^{-1}
\end{equation}
Define $B=kG$, $k^2=\lambda_1\lambda_2$, $l=k/\lambda_3$,$m=(\lambda_1+\lambda_2)k$ . Then eq. (\ref{b2}) can be rewritten as 
\begin{equation}
    G^2 =(m+l^{-1}) G-(ml^{-1}+1)I+l^{-1} G^{-1}
\end{equation}
Additionally, let us introduce another matrix
\begin{equation}
   E =(G+G^{-1})/m-I
\end{equation} 
BMW' can also be used for Yang-Baxterization, by writing the $\check{R}(u)$ matrix in the form:
\begin{equation}
    \check{R}(u)=I+ G(u)B+H(u)E+F(u)B
\end{equation}
Satisfying the BMW' algebra is a sufficient condition for Yang-Baxterizablity. Note that for different permutations of the eigenvalues, it is possible to obtain different solutions of the Yang-Baxter equation from the same BGR. This situation can be seen for $n\geq 3$.

\subsection{Example for Yang-Baxterization in $n=2$ case}

In this section, we consider an example of obtaining the Yang-Baxter equation from the Temperley-Lieb operators.

By putting $(I_i + f(u) E_i)$'s in the first inversion relation and using the relations of the TL algebra
\begingroup
\begin{align}
    &(I_i + f(u) E_i)(I_i + f(-u) E_i)=(I_j + f(u) E_j)(I_j + f(-u) E_j) \nonumber \\
    &=I_i + (f(u)+f(-u)) E_i + f(u)f(-u)E_i^2 \nonumber \\ &= I_j + (f(u)+f(-u)) E_j + f(u)f(-u)E_j^2 \nonumber \\
    &=I_i + (f(u)+f(-u)) E_i + f(u)f(-u) \delta E_i \nonumber \\ &=I_j + (f(u)+f(-u)) E_j + f(u)f(-u) \delta E_j
\end{align}
\endgroup
From here it can be seen that
\begingroup
\begin{align}
    &(f(u)+f(-u)+ \delta f(u)f(-u))E_i = (f(u)+f(-u)+ \delta f(u)f(-u))E_j \nonumber \\
    &(f(u)+f(-u)+ \delta f(u)f(-u))(E_i-E_j)=0
\end{align}
\endgroup

Since $E_i \neq E_j$ it has to be $f(u)+f(-u)+ \delta f(u)f(-u)=0$. Using this expression, one can easily find that
\begin{equation} \label{f(u)}
\frac{f(u)}{1 + \frac{\delta}{2}f(u)} =\frac{f(-u)}{1 + \frac{\delta}{2}f(-u)}
\end{equation}
From (\ref{f(u)}) it can be seen that $\frac{f(u)}{(1 +f(u))} =  \frac{2}{\delta} y(u)$ can be written where $y(u)$ is an odd function. Then $f(u)$ has the following form
\begin{align}
     f(u) = \frac{2}{\delta} \frac{y(u)}{1 - y(u)} 
\end{align}

Now we write the Yang-Baxter equation in terms of $\check{R}_i=I + f(u) E_i$ 
\begin{align}
& I + f(u)E_{i} +f(u+v)E_{i+1}+f(u)f(u+v)E_{i}E_{i+1}+f(v)E_{i} \\ & + f(u)f(v)E_{i}^2+f(u+v)f(v)E_{i+1}E_{i}+f(u)f(u+v)f(v)E_{i}E_{i+1}E_{i} \\
        & \qquad =I + f(v)E_{i+1} +f(u+v)E_{i}+f(v)f(u+v)E_{i+1}E_{i}+f(u)E_{i+1} \\ &+f(v)f(u)E_{i+1}^2+f(u+v)f(u)E_{i}E_{i+1}+f(v)f(u+v)f(u)E_{i+1}E_{i}E_{i+1}
\end{align}
and using the properties of $E_i$'s, one finds
\begin{align}
    &  f(u)E_{i} +f(u+v)E_{i+1}+f(v)E_{i}+f(u)f(v)\delta E_{i}+f(u)f(u+v)f(v)E_{i} \nonumber \\
    = & f(v)E_{i+1} +f(u+v)E_{i}+f(u)E_{i+1} +f(v)f(u)\delta E_{i+1} + f(v)f(u+v)f(u)E_{i+1}
\end{align}
Since $E_i - E_{i+1} \neq 0$ from the latter expression, we obtain the following result
\begin{align}
    &f(u)+f(v)+f(u)f(v)\delta + f(u)f(u+v)f(v)-f(u+v)=0 \;.
\end{align}
Writing $f(u)$'s in terms of $y(u)$'s, we can obtain a constraint on $y(u)$'s.
\begin{align} \label{exfory}
y(u)+y(v)-y(u+v)\left[ \left(1-\frac{4}{\delta^2}\right)y(u)y(v) \right] =0
\end{align} 
In the case $\delta = \pm 2$, we get $1-\frac{\delta^2}{4} = 0$ and eq. (\ref{exfory}) reduces to 
\begin{equation}
    y(u)+y(v)-y(u+v)=0 \;.
\end{equation}
An obvious solution to this equation is 
\begin{equation}
    y(u)=\frac{1}{k}u \;.
\end{equation}
Thus a rational solution has the following form
\begin{equation}
    f(u)=\pm \frac{u}{k-u} \;,
\end{equation}
where $k$ is an arbitrary constant that can be obtained. For the $\delta \neq \pm 2$ case there are other solutions that can be obtained \cite{li1993yang}. In the case $\delta = \pm 2$ or the case such that we can fix some parameters in $E$ such that $\delta = \pm 2$, we can obtain a rational solution of the Yang-Baxter equation
\begin{equation}
    \check{R}(u)=I\pm \frac{u}{k-u}E
\end{equation}
Now let us present an example of Yang-Baxterization which uses the mentioned way for obtaining rational solutions to the Yang-Baxter equation. Consider the braid group representation
\begin{equation}
    B = \begin{bmatrix}
    q & 0 & 0 & 0 \\
    t & 0 & p & 0 \\
    t & p & 0 & 0 \\
    \frac{2t^2}{q-p} & -t & -t & q
    \end{bmatrix}
\end{equation}
The eigenvalues of this matrix are $p$, $-p$, and $q$ where the multiplicity of $q$ is 2.

This matrix leads to two representations of the Temperley-Lieb algebra. One of them is 
\begin{equation}
    T = \begin{bmatrix}
        0 & 0 & 0 & 0 \\
        0 & -1 & 1 & 0 \\
        0 & 1 & -1 & 0 \\
        0 & 0 & 0 & 0
        \end{bmatrix}
\end{equation}
The $\sqrt{q}$ can be calculated in the relation $E_i^2=\sqrt{q}E_i$. 
\begin{equation}
    T^2 = \begin{bmatrix}
        0 & 0 & 0 & 0 \\
        0 & -1 & 1 & 0 \\
        0 & 1 & -1 & 0 \\
        0 & 0 & 0 & 0
        \end{bmatrix}^2 = 
        \begin{bmatrix}
        0 & 0 & 0 & 0 \\
        0 & 2 & -2 & 0 \\
        0 & -2 & 2 & 0 \\
        0 & 0 & 0 & 0
        \end{bmatrix} = -2T
\end{equation}
Using the recipe that has been stated previously, we can obtain 
\begin{equation}
    \check{R}(u)=I-\frac{u}{k-u}T = \begin{bmatrix}
        1 & 0 & 0 & 0 \\
        0 & \frac{k}{k-u} & \frac{-u}{k-u} & 0 \\
        0 & \frac{-u}{k-u} & \frac{k}{k-u} & 0 \\
        0 & 0 & 0 & 1
        \end{bmatrix}
\end{equation}

\section{Conclusion}
The $n$-CB algebras are a useful tool for Yang-Baxterizing BGR's. Yang-Baxter equation solutions are more general than BGR's, so they are often useful in finding link invariants. Using the relationship between braids and Yang-Baxter equation solutions, solutions of the Yang-Baxter equation can be used to construct link invariants. The computation of link invariants is complicated when using only the Boltzmann weights \cite{belavin20215}. However, the discussed $n$-CB algebras can be useful to simplify these calculations.

\section*{Acknowledgements} We would like to thank all the participants of the seminar series of ``Istanbul Integrability Initiative'' at Bogazici University. Ilmar Gahramanov is supported by the 3501-TUBITAK Career Development Program under grant number 122F451. The work of Ilmar Gahramanov is partially supported by the Russian Science Foundation grant number 22-72-10122. Cansu Özdemir wants to express her deepest gratitude to all of the ``Bir Nevi Akademi'' team, especially Hüseyin Ali Peker. CÖ is thankful for the support of Cayyolu Doga Science and Technology High School, especially Cavidan Varlı, Mahperi Karaca, Özlem Yüksel, and Gülay Aldemir. CÖ would like to thank Süreyya Topçu Ergin and Dr. Erkan Murat Türkan from Çankaya University. CÖ wants to express special thanks to Özgür Çıldıroğlu and Can Yürekli from Ankara University and her parents for their support. 

\bibliographystyle{utphys}
\bibliography{references}

\section{Appendix}
\subsection{Diagrams of the relations of 4-CB Algebra}

\begin{center}

\tikzset{every picture/.style={line width=0.75pt}} 



\end{center}

\subsection{Diagrams of the additional relations of 5-CB Algebra}

\begin{center}

\tikzset{every picture/.style={line width=0.75pt}} 

%

\end{center}

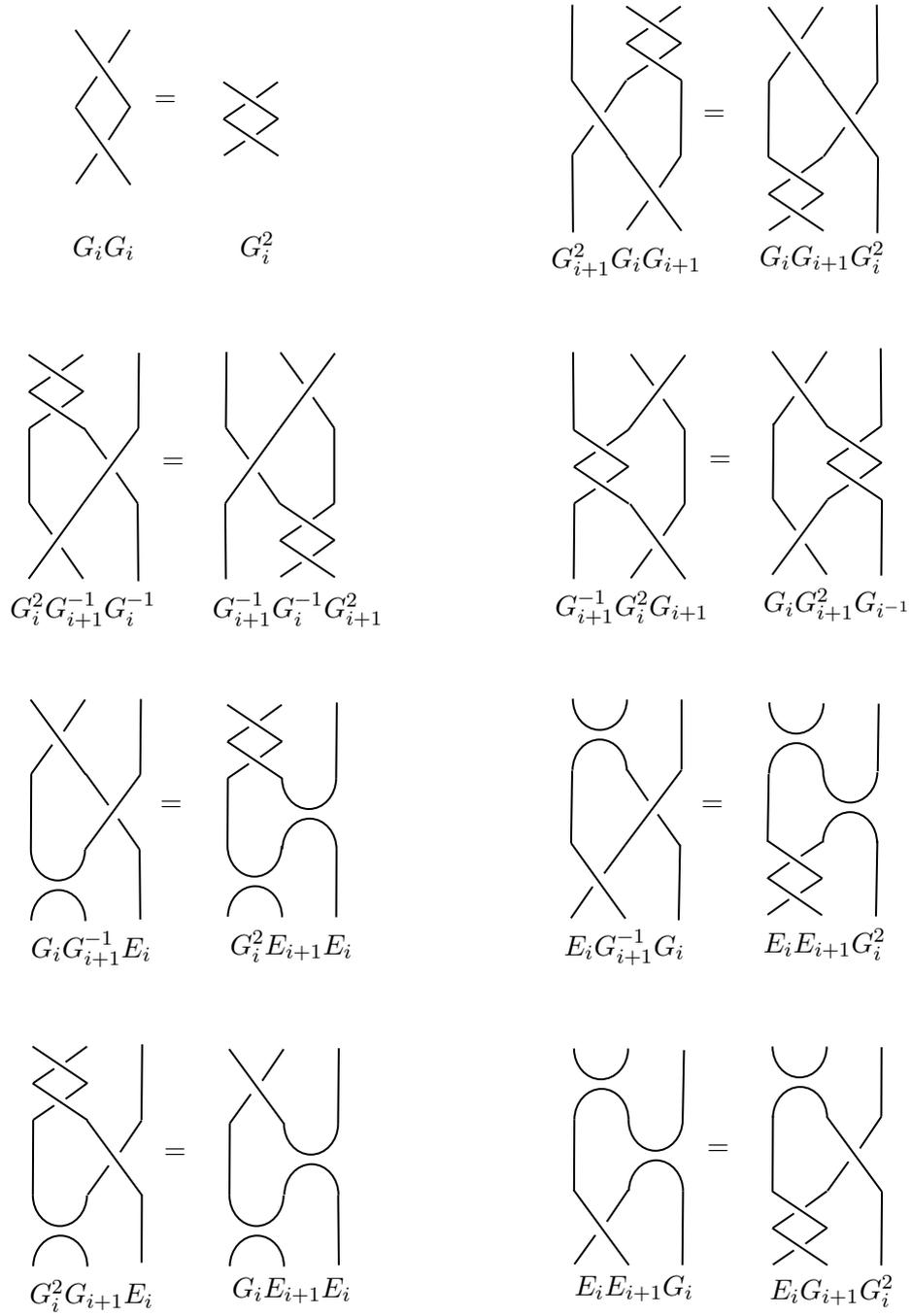
\end{document}